\documentclass[12pt]{article}
\usepackage[margin=1.2in]{geometry}
\usepackage{amsmath,amssymb,amsthm}
\usepackage[authoryear]{natbib}
\usepackage{mathtools,float}
\allowdisplaybreaks[4]
\usepackage[table,dvipsnames]{xcolor} 
\usepackage{booktabs}
\usepackage[T1]{fontenc}
\usepackage[latin9]{inputenc}
\usepackage{booktabs}
\usepackage{graphicx}
\usepackage{rotating}
\usepackage{comment}
\usepackage{wrapfig}
\usepackage{setspace}
\usepackage{titletoc}

\usepackage{subcaption}
\usepackage{float}

\definecolor{DarkBlue}{rgb}{0,0,0.6}
\usepackage{hyperref}
\hypersetup{
     colorlinks   = true,
     citecolor    = Maroon,
     urlcolor= DarkBlue,
     linkcolor=DarkBlue
}

\def\spacingset#1{\renewcommand{\baselinestretch}%
	{#1}\small\normalsize} \spacingset{1}
	
\graphicspath{ {./images/} }

\begin{document}

\title{\vspace{-1cm}Protectionism and economic growth:\\ Causal evidence from the first era of globalization\thanks{
We are grateful to Toke Aidt, Benoit Dicharry, Peter Egger, Carsten Eckel, Sibylle Lehmann-Hasemeyer, Mario Larch, Leif Lewin, Philippe Martin, Christian Merkl, Marc Muendler, Kevin O'Rourke, Panu Poutvaara, Laura Sabani, Claudia Steinwender, and the participants of the Silvaplana Political Economy Workshop 2020, the meeting of the Spanish Association of International Economics and Finance 2020, the meeting of the European Public Choice Society 2021, the 20th Journ\'ees Louis-Andr\'e G\'erard-Varet, the 2021 Annual Congress of the International Institute of Public Finance, the 2021 Annual Congress of the Verein f\"ur Socialpolitik and a seminar at the University of Cambridge for helpful comments, Viktor Persarvet and Henric H\"aggvist for sharing their data on the composition of Swedish imports and Swedish customs revenue, and Lukas Arth and Lukas K\"ahn for excellent research assistance. Fabian Ruthardt acknowledges funding from the Studienstiftung des deutschen Volkes - German Academic Scholarship Foundation - and the Konrad-Adenauer-Stiftung - Konrad-Adenauer-Foundation.}}

\author{Niklas Potrafke\thanks{Department of Economics, University of Munich and Ifo Institute, Ifo Center for Public Finance and Political Economy, Poschingerstr.\ 5, D-81679 Munich. Email: \url{potrafke@ifo.de}} \quad Fabian Ruthardt\thanks{Ifo Institute, Ifo Center for Public Finance and Political Economy, Poschingerstr.\ 5, D-81679 Munich. Email: \url{ruthardt@ifo.de}}\quad  Kaspar  W\"uthrich\thanks{Department of Economics, University of California San Diego, 9500 Gilman Dr.\ La Jolla, CA 92093; CESifo; Ifo Institute. Email: \url{kwuthrich@ucsd.edu}}}

\date{\today}

\maketitle

\spacingset{1} 

\vspace{-0.5cm}

\begin{abstract} 
We investigate how protectionist policies influence economic growth. Our empirical strategy exploits an extraordinary tax scandal that gave rise to an unexpected change of government in Sweden. A free-trade majority in parliament was overturned by a protectionist majority in 1887. The protectionist government increased tariffs. We employ the synthetic control method to select control countries against which economic growth in Sweden can be compared. We do not find evidence suggesting that protectionist policies influenced economic growth and examine channels why. The new tariff laws increased government revenue. However, the results do not suggest that the protectionist government stimulated the economy by increasing government expenditure.

\medskip



\end{abstract}

\newpage

\spacingset{1.4}


\section{Introduction}
How trade policies influence economic growth has been examined for a long time. Empirical evidence based on data for the late 20th and the early 21st century suggests that protectionist policies decrease economic growth.\footnote{See, for example,  \citet[][]{SachsWarner1995, Edwards1998, FrankelRomer1999, RodriguezRodrik2000, wacziarg2001measuring, irwintervio2002, vamvakidis2002, Dreher2006, wacziarg2008trade, nunntrefler2010, billmeier2013assessing, felbermayrgroeschl2013, eaton_etal2016, feyrer2019, Gygli2019, Irwin2019, andersen_etal2020, fajgelbaumetal2020, furcerietal2020}.} The empirical evidence from the late 19th and the early 20th century is less conclusive; most studies report positive correlations between tariffs and economic growth (`tariff-growth paradox').\footnote{See, for example, \citet[][]{bairoch1972EER, Irwin1998, Irwin2002, o'rourke2000, lehmann2011restat, schularick2011}. Similarly, \citet{pascali2017wind} reports that an increase in trade decreased economic growth for most countries in the first era of globalization.}
However, such positive correlations do not provide causal evidence on how protectionism influences growth because most policy changes are endogenous. Reverse causality and anticipation effects give rise to biases when applying, for example, standard panel data approaches based on international cross-sections. 

We provide causal evidence on the tariff-growth paradox by investigating a rare case of a plausibly exogenous change in trade policy. We exploit that an extraordinary tax scandal in the fall of 1887 gave rise to an unexpected change of government in Sweden. Swedish trade policies had been liberal for decades in the 19th century. Advocates of free trade (free-traders) also won the Swedish national elections in 1887 by a large margin. Two weeks after the election, a newspaper editor appealed the election results, claiming that a free-trade candidate from Stockholm is an illegitimate candidate because of outstanding tax liabilities. To the surprise of many, the election committee discarded all ballots with votes for the free-trade candidate and instated protectionist candidates as representatives for the Stockholm electoral district in the \textit{Riksdag}. In January 1888, the supreme court confirmed the decision of the election committee. The free-trade majority in the second chamber of parliament was overturned by a comfortable protectionist majority, and the free-trade government resigned. A protectionist government took office in February 1888 and increased tariffs by around 30\% \citep{Persarvet_2019}. 

The unanticipated change of government provides an ideal case for investigating how protectionist policies influence short-run economic growth.
First, because the change of government was unanticipated and decided by a court, anticipation effects and reverse causality are unlikely to bias our estimates.
Second, the tariff increase was large. Overall, tariffs increased by around 30\%, and all industries were affected by the tariff laws.
Third, customs revenue was the most important revenue stream for state finances at the time and made up 42\% of total government revenue in 1888/89 \citep{Haeggqvist2018}.
Fourth, trade policy was the central topic defining political competition, and the new tariff laws were the only major policy changes implemented by the protectionist government. Finally, changes in tariff laws were quickly perceived by merchants in the first era of globalization. The telegraph, for example, was used frequently in the late 19th century and reduced information frictions in international trade \citep{steinwender2018}.

We employ the synthetic control (SC) method \citep{abadie2003economic} to select control countries against which economic growth in Sweden can be compared. We do not find evidence suggesting that the protectionist policies influenced short-run economic growth in 19th century Sweden. The results corroborate that the short-run effects of protectionism are likely to be context-specific \citep{eichengreen2019}. Our study shows that focusing on exogenous variation is essential to better understand the `tariff-growth paradox' in the first era of globalization. 

A channel through which changes in import tariffs may influence short-run economic growth operates through changes in government revenue and fiscal policies. 
How changes in import tariffs affect imports depends on the tariff rate and the elasticity of import demand. We find that imports did not decrease and government revenue increased. Consequently, the protectionist government needed to decide how to spend the additional revenue. While increases in government expenditure translate quickly into higher GDP \citep{owyangetal2013, rameyzubairy2018, ramey2019}, our results do not show that the Swedish government increased government expenditure. The Swedish government used the additional revenue to consolidate budgets and repay public debt. Consistent with our empirical results, budget consolidation is unlikely to increase short-run economic growth.


Methodologically, our paper is most closely related to \citet{billmeier2013assessing} who employ the SC method to examine how trade policy reforms influence economic growth in the 20th century. Using the SC method in the literature on trade policies and economic growth has been a major innovation \citep{Irwin2019}. However, the variation in the trade policy reforms investigated by \citet{billmeier2013assessing} is not exogenous. While SC can accommodate some forms of selection on unobservables \citep[e.g.,][]{ferman2019properties}, it is unlikely to completely eliminate the endogeneity bias in such settings. We use the SC method and examine a change of government that induced exogeneous variation in Swedish trade policy.

Finally, we also contribute to the literature exploiting quasi-exogenous variation to examine how protectionism influenced economic development in the 19th century \citep[e.g.,][]{juhasz018}. We provide well-identified reduced-form evidence on the effect of protectionism on short-run growth in late 19th century Sweden.

\section{Change in government and protectionism}
\subsection{The 1887/1888 change in government}

Sweden pursued a liberal trade policy since the late 1850s \citep{Rustow1955}. In 1885, members of both chambers of the Swedish parliament started to organize themselves according to their stance on trade policy \citep{Rustow1955,Lewin1988}. The result was a face-off between free-traders and protectionists. The free-traders won the election in fall 1887 by a large margin \citep{Andersson1950}.\footnote{See \citet{LehmannVolckart2011} for a description of the electorates of free-traders and protectionists.} Thus, it was very likely that the liberal trade policy would have been continued. 

Shortly after the fall election, an unexpected event took place, which was called ``sensational'' \citep{Lewin1988}, ``preposterous'' \citep{CarlssonRosen1991}, and ``scandalous'' \citep{esaiasson1990svenska}.
Stockholm's electoral district was entitled to 22 seats in the second chamber of parliament \citep{Rustow1955}. Citizens in Stockholm elected only free-traders into parliament by large vote margins.\footnote{Stockholm was the main stronghold of free-trade sentiment at the time. See Appendix \ref{sec: Election results Stockholm} for the fall election results for the electoral district of Stockholm.} 
The election's appeal period lasted until October 4, 1887. Two citizens filed appeals against the election results in Stockholm's electoral district \citep{stockholmsdagblad_18871005}. The appeal by Wilhelm Alexander Bergstrand, the publisher of the newspaper \textit{Nya Dagligt Allehanda}, induced political turmoil in Stockholm and soon after in the whole country.

On October 4, 1887, shortly before the appeal period ended, Bergstrand submitted his appeal and published it in \textit{Nya Dagligt Allehanda} on the same day \citep{NyaDagligtAllehanda_18871004}. In his appeal, Bergstrand claimed that Olof Larsson, one of the 22 free-trade candidates, owed a small amount of crown and municipal taxes for 1881 and 1882. According to paragraphs 25 and 26 of the Parliament Act of 1866, a candidate with tax debt is disqualified, and all ballots with votes for the respective candidate are invalid \citep{Riksdagsordningen1866}. Bergstrand demanded that all ballots with votes for Larsson must be declared invalid. He further demanded a recount of all valid votes. On October 5, 1887, Bergstrand published proof for Larsson's tax liabilities: the tax collection commissioner for Adolf Fredriks and Kungsholms (two districts in Stockholm) had issued a certificate confirming Larsson's tax liabilities on October 4, 1887 \citep{NyaDagligtAllehanda_18871005}.

\begin{figure}[H]
	\caption{The 1887/1888 change in government and the 1890 general election}
	\begin{center}
		\includegraphics[width=0.96\textwidth]{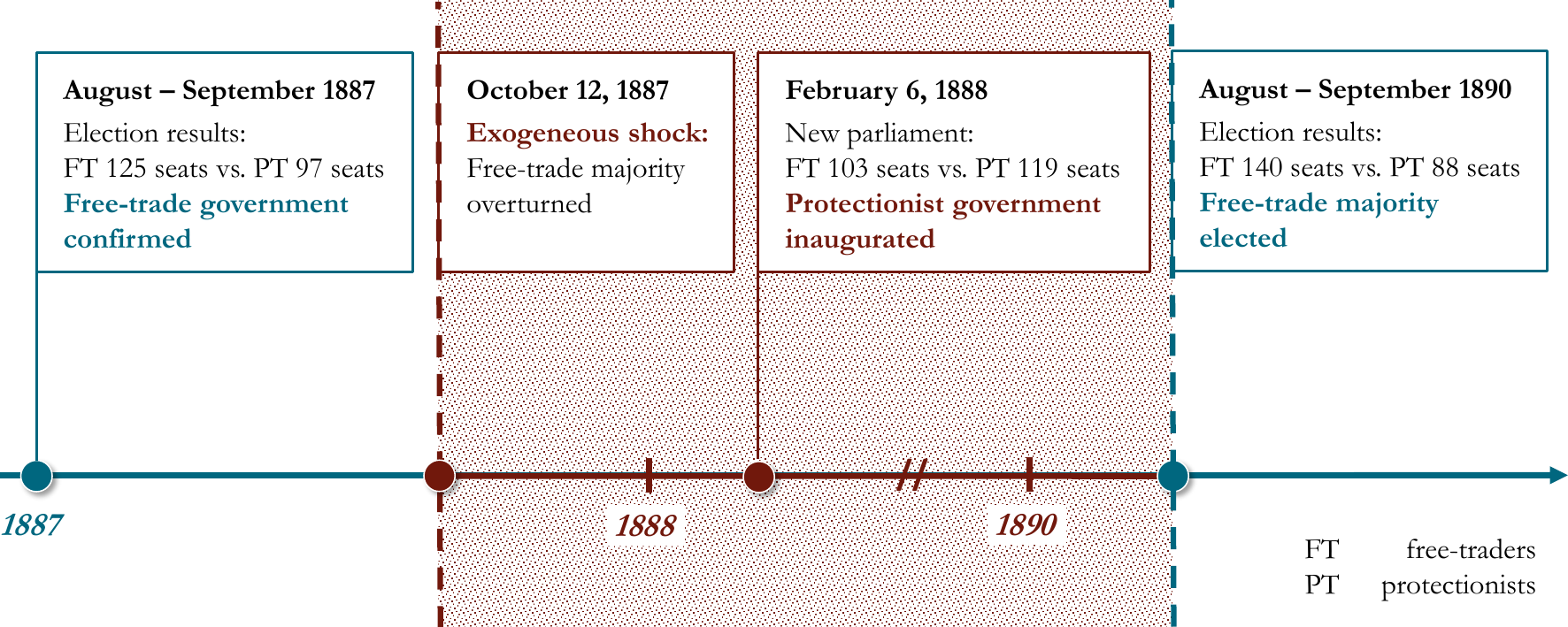}
	\end{center}
	\footnotesize{\textit{Source:} Own illustration}
	\label{fig: The 1887/1888 change in government and the 1890 general election}
\end{figure}

Events unraveled during the following days. Many newspapers published opinions about the legitimacy of the appeal. Larsson's statement in \textit{Aftonbladet}, one of the most influential newspapers at the time, disputed any tax liabilities but remained without the intended effect \citep{Aftonbladet_18871007}. On October 12, 1887, the election committee accepted Bergstrand's appeal and invalidated all ballot papers with votes for Larsson \citep{Lindorm1936}. It ordered a recount of the votes and declared the 22 protectionist candidates winners of the election. Disputes followed and the decision of the election committee was challenged. On January 25, 1888, the supreme court ruled that the 6,585 ballot papers with votes for Larsson are indeed invalid and officially instated the 22 protectionist candidates as legitimate representatives of the electoral district of Stockholm in the \textit{Riksdag} \citep{Lewin1988}. The free-trade majority in the second chamber of parliament (125 free-traders, 97 protectionists) was overturned by a comfortable protectionist majority (119 protectionists, 103 free-traders).\footnote{Both chambers of parliament decide on trade policy, and each representative has one vote.} As a result, the free-trade government resigned on February 6, 1888, and the experienced protectionist Gillis Bildt became Prime Minister \citep{Lindorm1936}.\footnote{Bildt served as Swedish ambassador in Berlin when the \textit{Reichstag} under Bismarck introduced the agrarian protectionist system in 1879.}
In February 1888, Bildt's government issued the first tariff laws. See Section \ref{sec:tariffs} for details on the new tariff laws.  Figure \ref{fig: The 1887/1888 change in government and the 1890 general election} shows the timeline of the main events.

The change in parliamentary majorities in the aftermath of the 1887 fall election occurred unexpectedly. We reviewed hundreds of articles from regional and national Swedish newspapers from before the September 1887 election up to January 25, 1888.\footnote{We used a search algorithm with keywords and time periods for Swedish newspaper articles provided by the National Library of Sweden (\textit{Kungliga biblioteket, KB}).} We found no indication that the tax debt was publicly known before the election.

\subsection{Swedish protectionist policies}
\label{sec:tariffs}
The protectionist government increased overall tariffs by around 30\% in 1888 \citep{Persarvet_2019}.\footnote{We contribute to the long-standing debate on the effect of the Swedish tariff increases by providing causal evidence \citep[e.g.,][]{heckscher1941svenskt,montgomery1966industrialismens,jorberg1961growth,jorberg1966naagra,hammarstrom1970stockholm,schon1989kapitalimport,Bohlin2005,Haeggqvist2018,Persarvet_2019}. We review this debate in more detail in Appendix \ref{sec: The Swedish tariff debate}.} The tariff increase was heterogeneous across product classes.

We follow \citet{Persarvet_2019} and classify the goods of the Swedish historical trade statistics according to the Standard International Trade Classification (SITC) framework. Tariffs on food and beverages increased substantially (SITC sections 0--1). The protectionist government raised food tariffs on average by six percentage points. The increase affected 36\% of total imports. The largest tariff increase was on grain (from 2\% to 27\%).

Tariffs on raw materials and fuels increased only to a small extent (SITC sections 2--4). The tariff increase on scrap metal increased the average tariff on ores and metal scrap. New tariffs on lard increased the average tariff on animal and vegetable fats. Coal, coke, and crude oil remained duty-free.

Tariffs on manufactured products increased slightly (SITC sections 5--9). Most of the industrial tariffs were still bound by the Franco-Swedish trade agreement.\footnote{In 1860, France and Great Britain signed the Cobden-Chevalier treaty. This triggered a large number of most favored nation treaties on the European continent and contributed to a period of relatively free trade \citep{Lampe2009,tena2012}. France and Sweden signed a trade agreement in 1865. When this agreement expired in 1892, Sweden regained tariff autonomy and substantially increased tariffs on industrial products \citep{Persarvet_2019}.} Tariffs increased on iron and steel products through the introduction of new tariffs on sheet metal, steel beams, cast steel, and metal wire.

\section{Data and empirical strategy}

\subsection{Data}
We use data from the Jord\`{a}-Schularick-Taylor Macrohistory Database \citep{JordaSchularickTaylor2017}.\footnote{The data are available here: \url{http://www.macrohistory.net/data/}.}
The Jord\`{a}-Schularick-Taylor Database includes annual data for 17 advanced economies since 1870. It encompasses measures of GDP\footnote{We use real GDP per capita (index, 2005=100).}, imports, central government revenue, and central government expenditure. Data comes from a broad range of historical sources and various publications of governments, statistical offices, central banks, and private banks. For some countries, the authors extended data series from university databases and international organizations. The main source for our GDP measure is the Macroeconomic Data Set \citep{BarroUrsua2010}. Most trade and national account data come from \citet{mitchel2007international}, \citet{flora1983state}, IMF international financial statistics, OECD national accounts statistics, and national statistics offices.

We examine data until 1890 because the next election took place in the fall of 1890. The free-traders won this election. An important advantage of using a relatively short post-treatment period is that other potential confounding events are unlikely to affect our analysis. 

\subsection{The synthetic control method}

We employ the SC method \citep{abadie2003economic,abadie2010synthetic,abadie2015comparative}; see \citet{abadie2020jel} for a review.\footnote{There is a growing body of work using SC to make causal inference in aggregate panel data settings \citep[e.g.,][]{billmeier2013assessing,bohn2014did,pinotti2015economic,cunningham2018decriminalizing, asatryan_etal2018, eliason2018can, andersson2019carbon,born2019costs,powu2020}.} SC approximates what would have happened to Sweden with a free-trade government using a weighted average of control countries. We perform the empirical analyses in \texttt{Stata} \citep{stata2019} and \texttt{R} \citep{R2020}.

To describe the SC method formally, we use the potential outcomes framework \citep{rubin1974estimating}. We denote by $Y_{jt}^{F}$ and $Y_{jt}^{P}$ the potential outcome of country $j$ in period $t$ with a free-trade ($F$) and a protectionist ($P$) government. Our main outcome of interest is real GDP per capita, and we also investigate imports, government revenue, and government expenditure.
Let $j=1$ index Sweden and $j=2,\dots,J+1$ index the $J$ control countries. We discuss the choice of the $J$ control countries, our \emph{donor pool}, in Section \ref{sec:donor_pool}.

Our purpose is to estimate the causal effect of protectionism between 1888 and 1890 (the year of the next election),
\begin{equation}
    \tau_t=Y_{1t}^{P}-Y_{1t}^{F}, \quad t\in\{1888,1889,1890\}.
\end{equation}
For Sweden, we observe $Y_{1t}^{F}$ until 1887 and $Y_{1t}^{P}$ afterwards. For the control countries, we observe $Y_{jt}^{F}$ for all periods. Thus, to estimate $\tau_t$, we need to estimate the unobserved potential outcome $Y_{1t}^{F}$. We use the following estimator 
\begin{equation}
    \hat{Y}_{1t}^F=\sum_{j=2}^{J+1}\hat{w}_jY_{jt}^F.\label{eq:sc}
\end{equation}
We refer to the weighted average in equation \eqref{eq:sc} as \emph{synthetic Sweden}.
The SC weights $\left(\hat{w}_2,\dots,\hat{w}_{J+1}\right)$ are estimated by minimizing the discrepancy between the pre-treatment outcomes for Sweden and synthetic Sweden using the Stata package \texttt{synth} \citep{synth}. To avoid concerns about specification search, we do not include additional predictors. The weights are restricted to be positive and add up to one, which ensures transparency and precludes extrapolation \citep[][Section 4]{abadie2020jel}. 

SC generalizes difference-in-differences (DID). To approximate $Y_{1t}^F$, DID employs simple averages of control units chosen by the researcher. By contrast, SC chooses controls in an automatic data-driven way, employing a weighted average (equation \eqref{eq:sc}) to approximate $Y_{1t}^F$. As a result, SC is less susceptible to specification search and often provides a better counterfactual approximation. See Section 4 in \citet{abadie2020jel} for further discussions of the advantages of SC.

To make inferences, we employ the permutation method proposed by \citet{abadie2010synthetic}. See \citet{firpo18synthetic} and \citet{abadie2020jel} for further discussions. In Section \ref{sec:robustness_conformal}, we apply the conformal inference procedure of \citet{chernozhukov2020exact} as an additional robustness check.\footnote{We implement the conformal inference procedure using the \texttt{R}-package \texttt{scinference} (available here: \url{https://github.com/kwuthrich/scinference}).}

\subsection{Choice of donor pool}
\label{sec:donor_pool}
We restrict our donor pool of control units to countries that had free-trade governments from 1870 to 1890. From the 17 countries available in the Jord\`{a}-Schularick-Taylor Database, we exclude France, Germany, Italy, Spain, and Portugal because of protectionist trade policies.\footnote{We use country classifications of previous studies \citep[e.g.,][]{O'Rourke1996, o'rourke2000, Irwin1998, Irwin2002, RodriguezRodrik2000, ClemensWilliamson2004, Williamson2006, schularick2011} and classify countries either as ``protectionist''/``tariff hikers'' or ``free-trade''/``non-tariff hikers''.} Data is missing for Australia and Japan. Therefore, our donor pool includes Belgium, Canada, Denmark, Finland, the Netherlands, Norway, Switzerland, the United Kingdom, and the United States. In Section \ref{sec:robustness_conformal}, we present results for a restricted donor pool with only European countries.

An important requirement for SC analyses is that the donor pool of control countries is homogeneous enough \citep{abadie2020jel}. All countries in our donor pool were industrializing during the 1870s and 1880s. Citizens or elected representatives of the citizens possessed substantial political power and influenced national policies.

\section{Results: protectionism and growth}

\begin{figure}[H]
	\caption{Real GDP per capita}
	\begin{center}
		\includegraphics[width=0.48\textwidth]{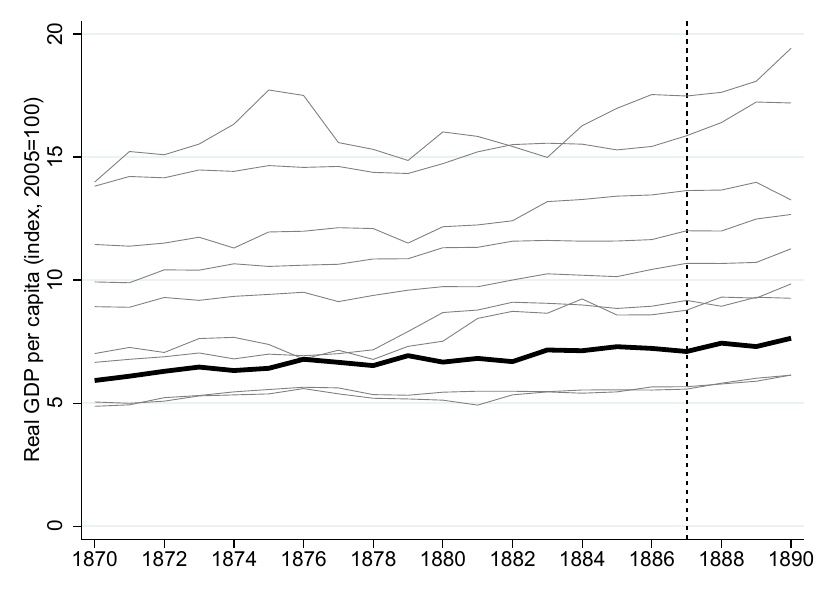}
		\includegraphics[width=0.48\textwidth]{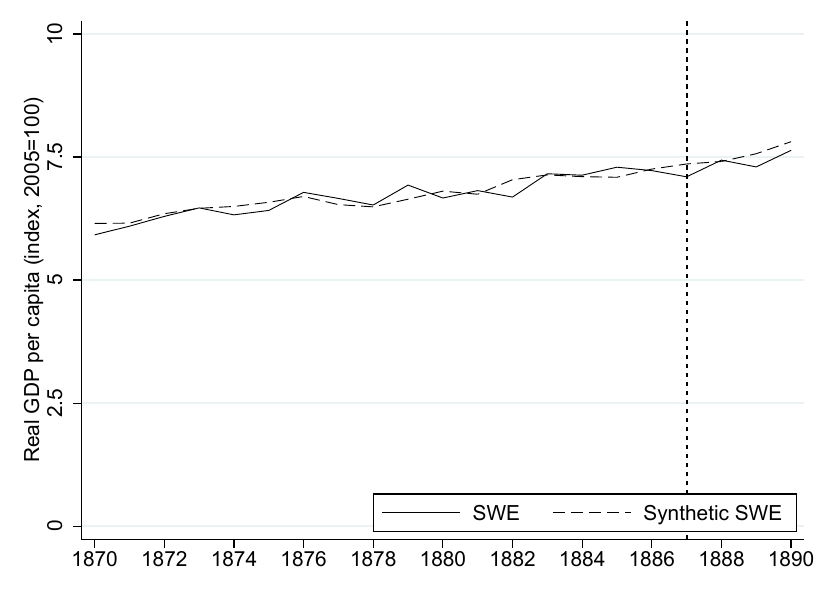}
        \includegraphics[width=0.48\textwidth]{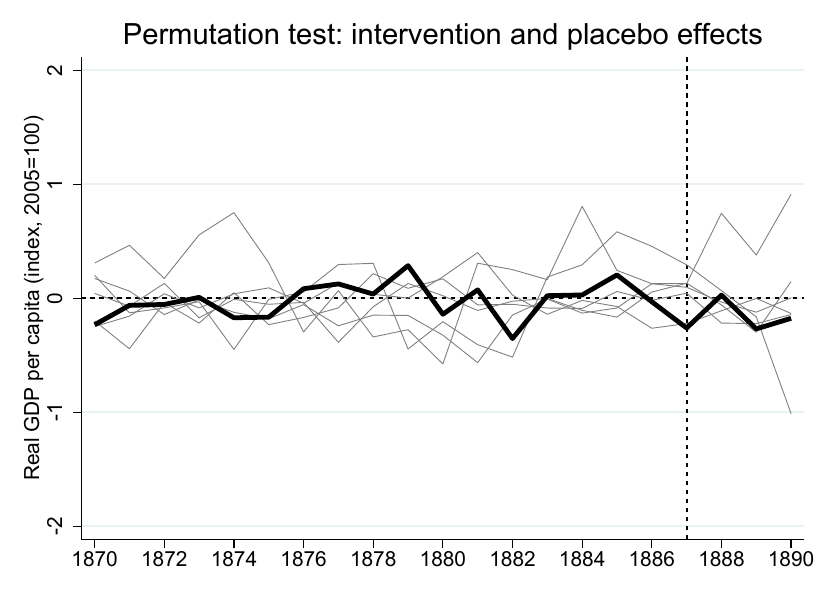}
        \includegraphics[width=0.48\textwidth]{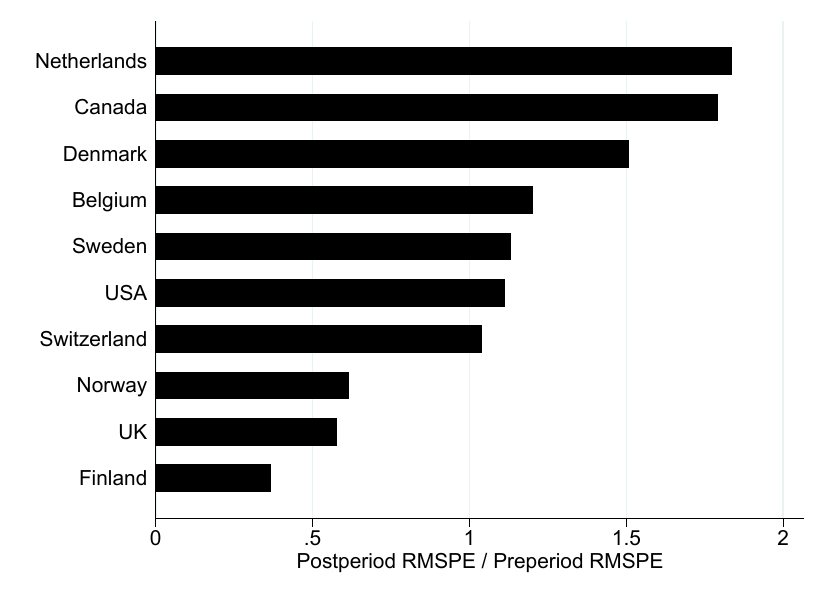}
	\end{center}
	\footnotesize{\textit{Notes:} Real GDP per capita is shown as an index (2005 = 100). The lower left panel excludes countries for which the pre-treatment MSPE is at least 10 times larger than Sweden's pre-treatment MSPE. The data are from the Jord\`{a}-Schularick-Taylor Macrohistory Database.}
	\label{fig: Real GDP per capita}
\end{figure}

The upper left panel of Figure \ref{fig: Real GDP per capita} shows real GDP per capita for each donor pool country and Sweden from 1870 to 1890. Sweden's GDP is shown in thick black; the control countries' GDPs are shown in grey. The upper right panel shows how real GDP per capita developed in Sweden and synthetic Sweden over the period 1870--1890. The synthetic Sweden consists of 21.7\% Denmark, 43.6\% Finland, 17.3\% Norway, 0.3\% United Kingdom, and 17.0\% United States (Table \ref{tab: Weights} in Appendix \ref{sec: Weights}).

We find no evidence suggesting that protectionism influenced real GDP per capita. From 1870 to 1887, Sweden's average real GDP per capita grew from 5.92 to 7.10 (average annual growth rate (AAGR): 1.07\%), and synthetic Sweden's average real GDP per capita grew from 6.15 to 7.36 (AAGR: 1.06\%). After the change in government, from 1887 to 1890, Sweden's average real GDP per capita grew from 7.10 to 7.64 (AAGR: 2.47\%), and synthetic Sweden's average real GDP per capita grew from 7.36 to 7.81 (AAGR: 2.01\%).  

To make inferences, following \citet{abadie2010synthetic}, we iteratively re-assign the treatment of having a protectionist government to each country in the donor pool. Because SC does not yield good pre-treatment fits for some control countries, we exclude countries for which the pre-treatment mean squared prediction error (MSPE) is more than 10 times larger than the pre-treatment MSPE for Sweden (lower left panel of Figure \ref{fig: Real GDP per capita}). The results do not suggest that the effect of protectionism on GDP in Sweden was large relative to the distribution of placebo effects. Since the cutoff of 10 is arbitrary, we also report the ratios of post-treatment root MSPE (RMSPE) to pre-treatment RMSPE, as suggested by \citet{abadie2015comparative}. A large RMSPE ratio indicates a rejection of the null hypothesis that protectionism had no effect. The lower right panel of Figure \ref{fig: Real GDP per capita} suggests that Sweden's ratio was not large compared to the other countries in the donor pool.

In Section \ref{sec:robustness_conformal}, we show that our results are robust to potential spillover effects from Sweden's tariff policy on its trading partners and to restricting the donor pool to European countries.

\section{Channels}
We examine channels for why there is no evidence suggesting that protectionism influenced short-run economic growth. We focus on outcomes of international trade and fiscal policies that are likely to influence short-run economic growth.\footnote{Clearly, if possible, we would have liked to also examine the extent to which short-run economic growth was influenced through changes in firms' productivity. However, there is no data on firms' productivity or TFP across countries in the late 19th century available, which we could use in our SC analyses.}

\subsection{Imports}
\label{sec:imports}
\begin{figure}[H]
	\caption{Imports as a share of GDP}
	\begin{center}
		\includegraphics[width=0.48\textwidth]{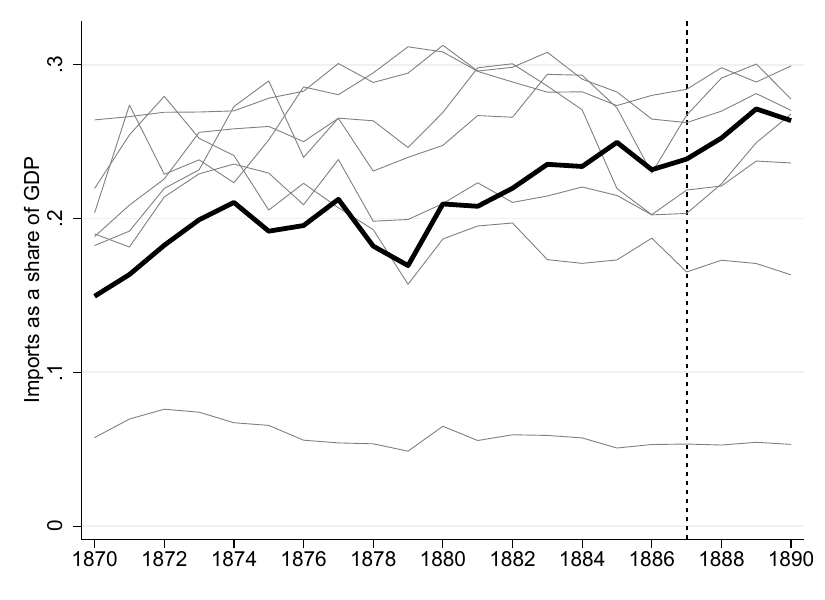}
		\includegraphics[width=0.48\textwidth]{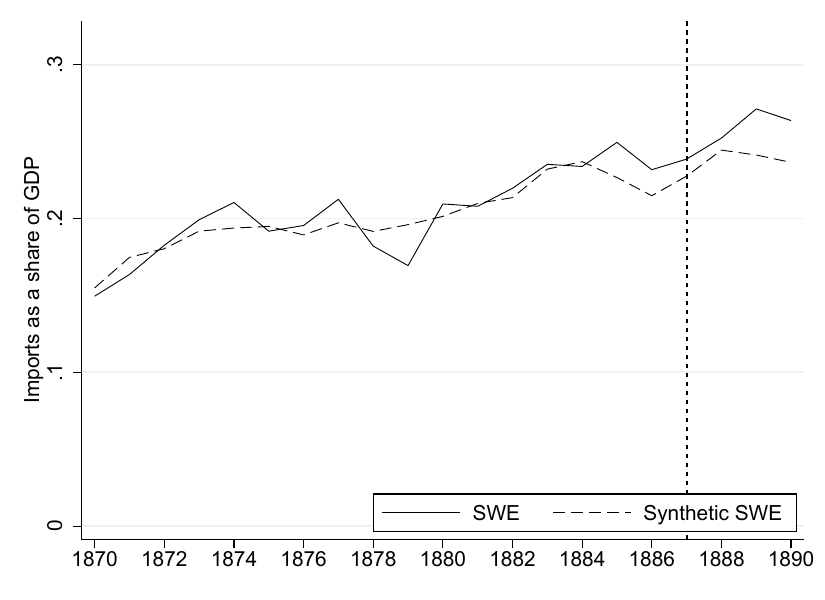}
       \includegraphics[width=0.48\textwidth]{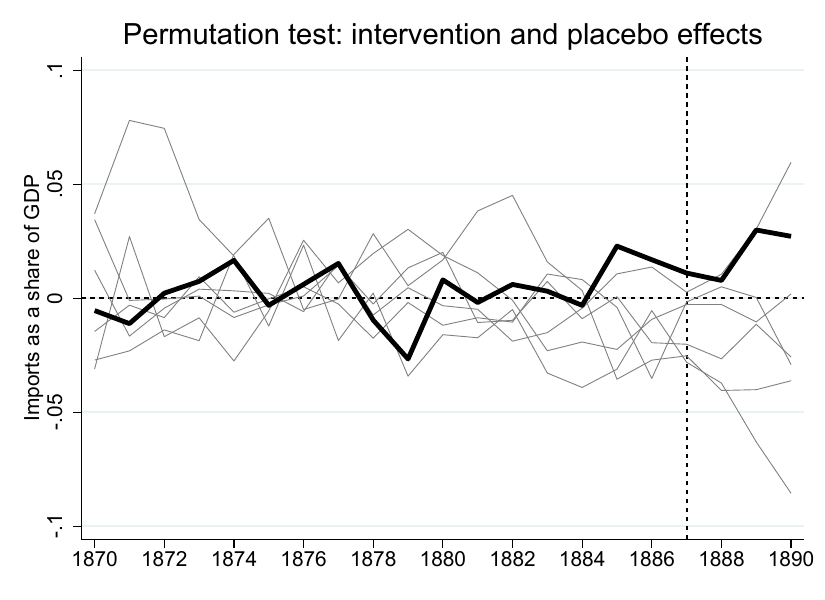}
        \includegraphics[width=0.48\textwidth]{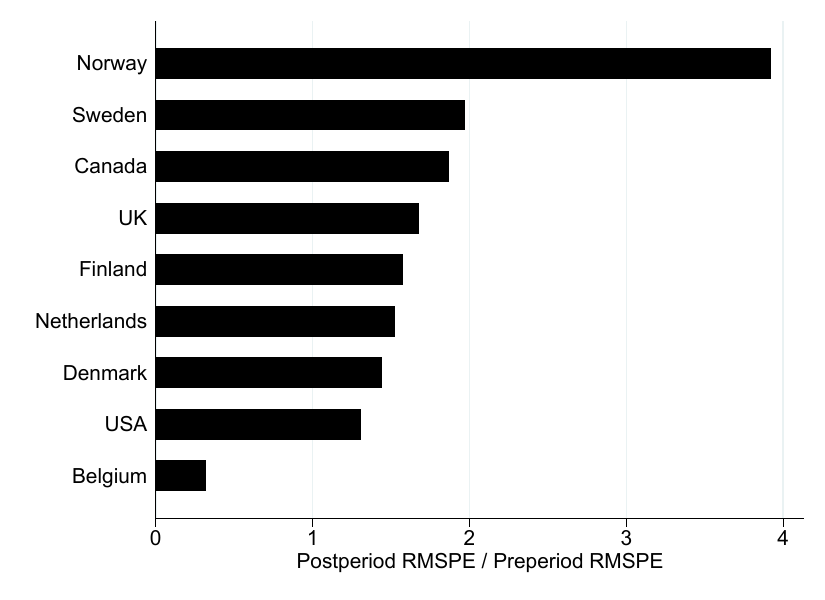}
	\end{center}
	\footnotesize{\textit{Notes:} The graphs (upper left panel) do not include the Netherlands' imports. The Netherlands' imports as a share of GDP were between 54.43\% and 107.95\% (1870--1890). Data is missing for Switzerland. The lower left panel excludes countries for which the pre-treatment MSPE is at least 10 times larger than Sweden's pre-treatment MSPE. The data are from the Jord\`{a}-Schularick-Taylor Macrohistory Database.}
	\label{fig: Imports as a Share of GDP}
\end{figure}

It is conceivable that protectionism decreased imports, especially from those countries from which Sweden imported a substantial fraction of its goods. However, the empirical results in Figure \ref{fig: Imports as a Share of GDP} do not suggest that the introduction of tariffs decreased imports. The total value of imports increased from 297.41 million SEK in 1887 to 324.71 million SEK in 1888. 
The protectionist tariff policy implemented in early 1888 did not reverse the steady growth of imports. The total value of imports as a share of GDP increased from 14.95\% in 1870 to 23.87\% in 1887. In 1888, imports as a share of GDP increased to 25.23\% and reached 26.37\% in 1890. 

We do not find evidence that aggregate import levels masked heterogeneous effects of the Swedish tariffs on individual trading partners. See Appendix \ref{sec: Imports by trade partner} for how Swedish imports from individual countries developed between 1870 and 1890.

Bildt's government increased tariffs to different extents across sectors (see Section \ref{sec:tariffs}). Appendix \ref{sec: Swedish imports by sector} shows how the composition of Swedish imports across sectors developed between 1870--1890. Agricultural imports remained stable on a high level, and manufactured imports continued their growth path after 1888. Based on our data, we cannot disentangle the effects of tariffs on agricultural imports and tariffs on manufactured imports. However, we do not observe that the composition of imports changed substantially after 1887.

\subsection{Government revenue}
\label{sec: govt revenue}

We examine how the protectionist policies influenced government revenue. Whether the higher tariffs decreased or increased government revenue depends on the tariff rate and the elasticity of import demand. As shown in Section \ref{sec:imports}, imports did not decrease when tariffs were increased. Hence, higher tariffs may well have increased government revenue. 

\begin{figure}[H]
	\caption{Government revenue as a share of GDP}
	\begin{center}
		\includegraphics[width=0.48\textwidth]{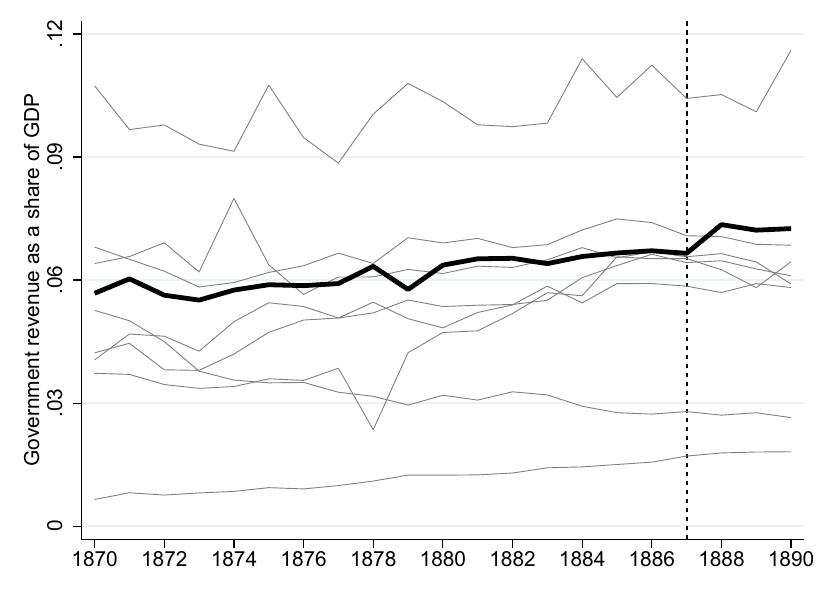}
		\includegraphics[width=0.48\textwidth]{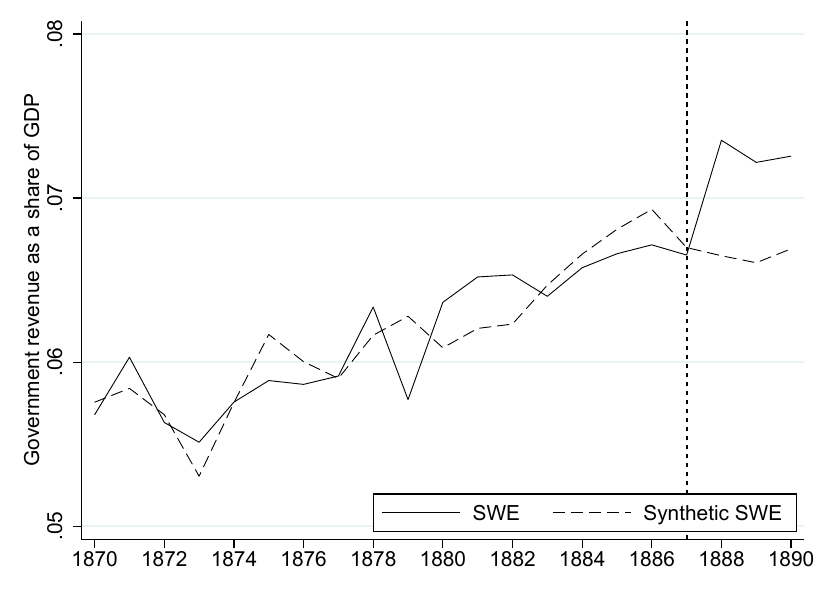}
        \includegraphics[width=0.48\textwidth]{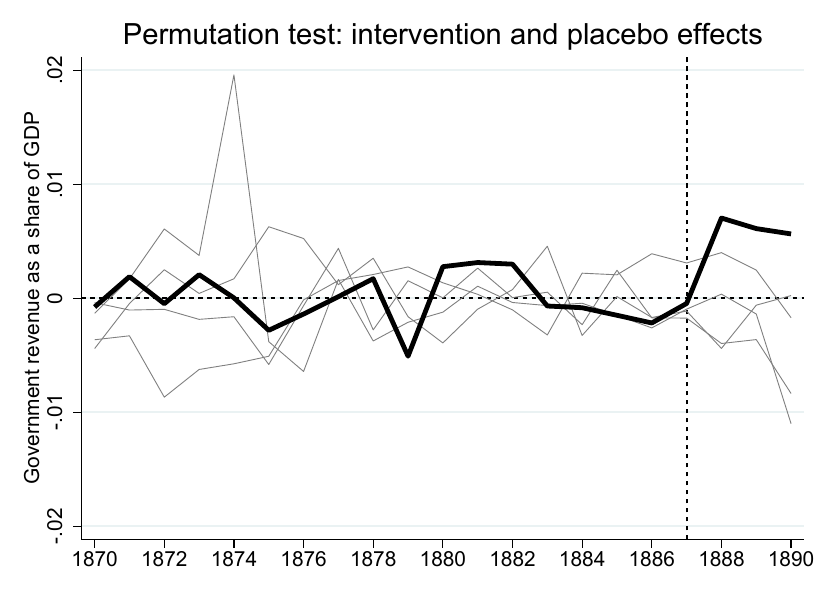}
        \includegraphics[width=0.48\textwidth]{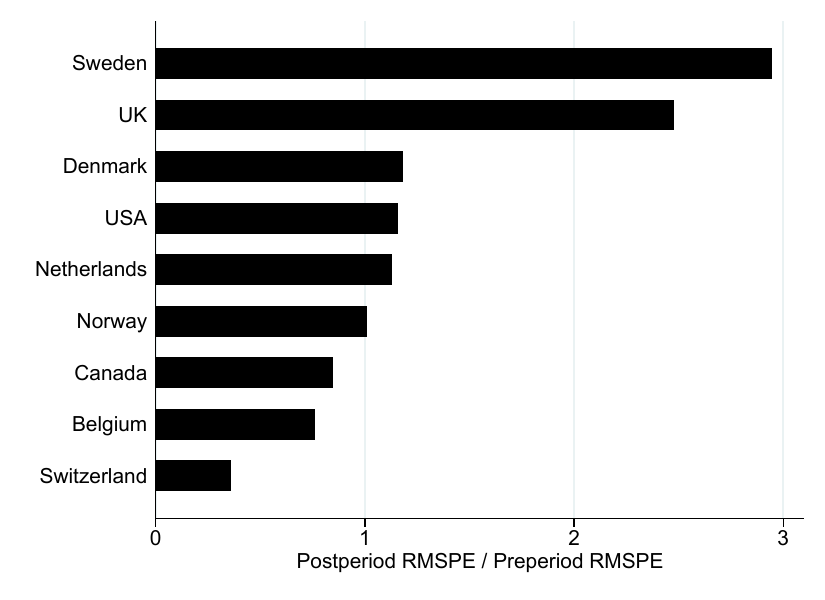}
	\end{center}
	\footnotesize{\textit{Notes:} Data is missing for Finland. The lower left panel excludes countries for which the pre-treatment MSPE is at least 10 times larger than Sweden's pre-treatment MSPE. The data are from the Jord\`{a}-Schularick-Taylor Macrohistory Database.}
	\label{fig: Government Revenue as a Share of GDP}
\end{figure}

Figure \ref{fig: Government Revenue as a Share of GDP} shows that the protectionist policies enacted after the change of government increased government revenue. The ratio of the post-treatment to the pre-treatment RMSPE is the largest for Sweden. If one were to select a country at random, the probability of obtaining a ratio as high as Sweden's is $1/9$ \citep[see][for a further discussion of this interpretation]{abadie2015comparative}. This result is robust to using the conformal inference procedure of \citet{chernozhukov2020exact} (cf.\ Section \ref{sec:robustness_conformal}).

Government revenue was 81.11 million SEK in 1887. It increased by 16.02\% to 94.11 million SEK in 1888. As a share of GDP, government revenue increased from 6.65\% to 7.35\% and remained relatively stable until 1890 (1889: 7.22\%, 1890: 7.26\%). 
Meanwhile, synthetic Sweden's government revenue as a share of GDP decreased from 6.70\% in 1887 to 6.65\% in 1888. It remained relatively stable until 1890 (1889: 6.61\%, 1890: 6.69\%). Customs revenue was responsible for the increase in government revenue (see Appendix \ref{sec: Swedish fiscal policies in the first era of globalization} for a description of Swedish fiscal policies 1888--1890). Because imports did not decrease when the protectionist policies were introduced, it is unlikely that tariffs were systematically circumvented.\footnote{Further, given the development of Swedish imports from Norway after 1887, it is unlikely that goods destined for Sweden were shipped to Norway and then crossed country borders on rail; see Appendix \ref{sec: Imports by trade partner}.}

Our results are in line with empirical evidence from the United States in the 1880s \citep{Irwin1998}. The United States enjoyed high fiscal surpluses in the early 1880s, and the political parties were discussing how changing import tariffs would influence customs revenues. \citet{Irwin1998} estimates the revenue effects of the proposed tariff changes and concludes that lower import tariffs would also have reduced customs revenues. 

\subsection{Government expenditure}
\label{sec: Government expenditure}

\begin{figure}[H]
	\caption{Government expenditure as a share of GDP}
	\begin{center}
		\includegraphics[width=0.48\textwidth]{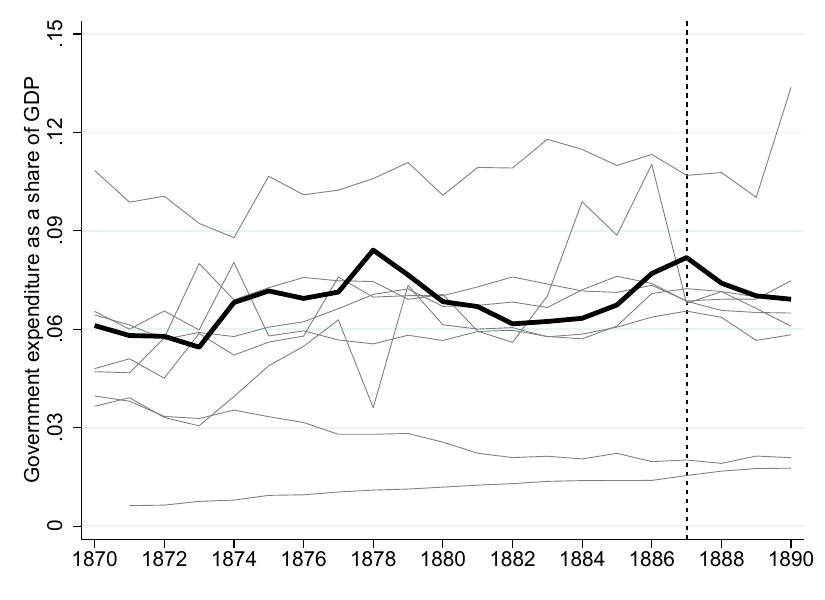}
		\includegraphics[width=0.48\textwidth]{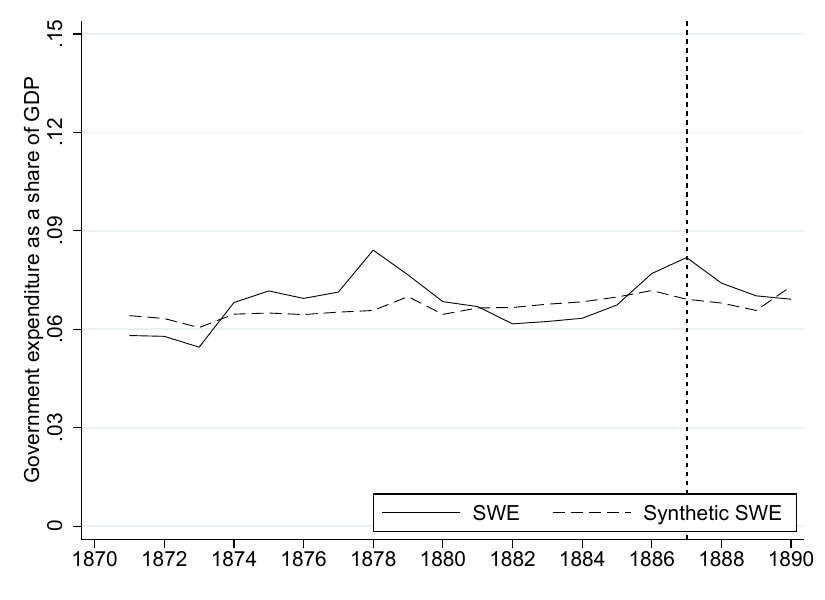}
       \includegraphics[width=0.48\textwidth]{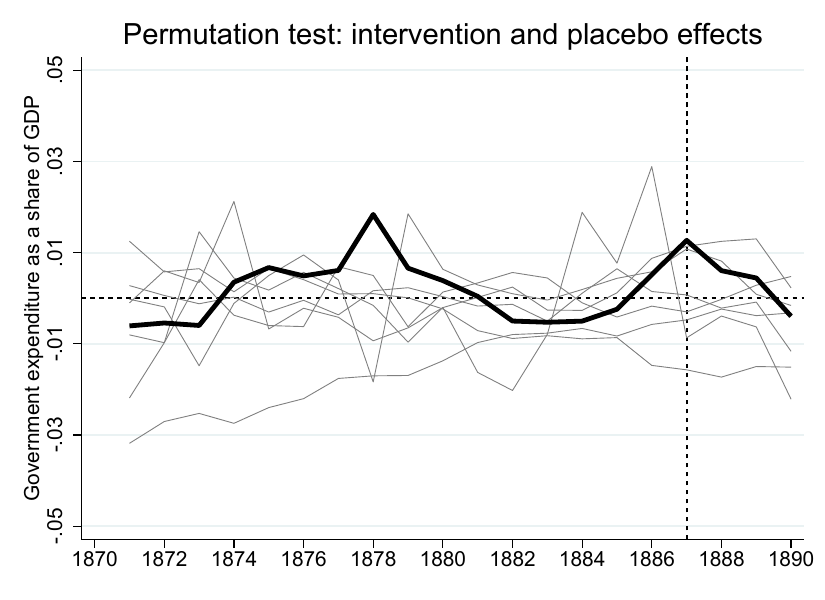}
        \includegraphics[width=0.48\textwidth]{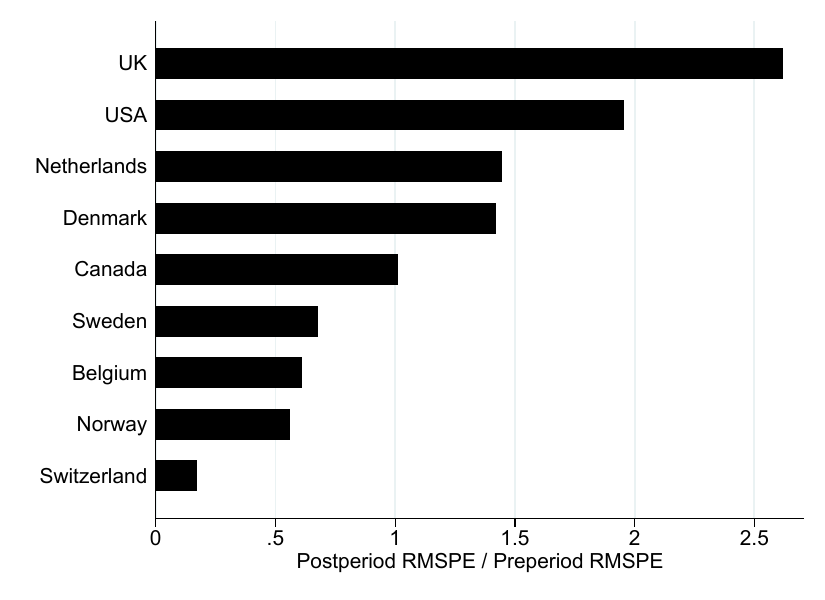}
	\end{center}
	\footnotesize{\textit{Notes:} Data is missing for Switzerland's government expenditure in 1870. Therefore, we calculate the synthetic Sweden based on the best pre-treatment fit from 1871 to 1887. Data is missing for Finland. The lower left panel excludes countries for which the pre-treatment MSPE is at least 10 times larger than Sweden's pre-treatment MSPE. The data are from the Jord\`{a}-Schularick-Taylor Macrohistory Database.}
	\label{fig: Government expenditure as a Share of GDP}
\end{figure}

Figure \ref{fig: Government expenditure as a Share of GDP} shows the SC estimates for government expenditure. The results do not suggest that the protectionist government influenced government expenditure. Swedish government expenditure as a share of GDP decreased from 8.19\% in 1887 to 6.92\% in 1890. Synthetic Sweden's government expenditure as a share of GDP increased from 6.92\% to 7.31\% over the same period.

Sweden went from a large primary budget deficit in 1887 to a small primary budget surplus in 1888. The surplus increased in 1889 and 1890. In both years, Sweden had a total budget surplus and total government debt decreased. 

\section{Robustness checks}
\label{sec:robustness_conformal}
We submit the estimated effect of protectionism on government revenue to four robustness checks and also apply the conformal inference procedure of \citet{chernozhukov2020exact}. 

\begin{figure}[H]
	\caption{Placebo treatment in 1884 and leave-one-out sensitivity}
	\begin{center}
        \includegraphics[width=0.48\textwidth]{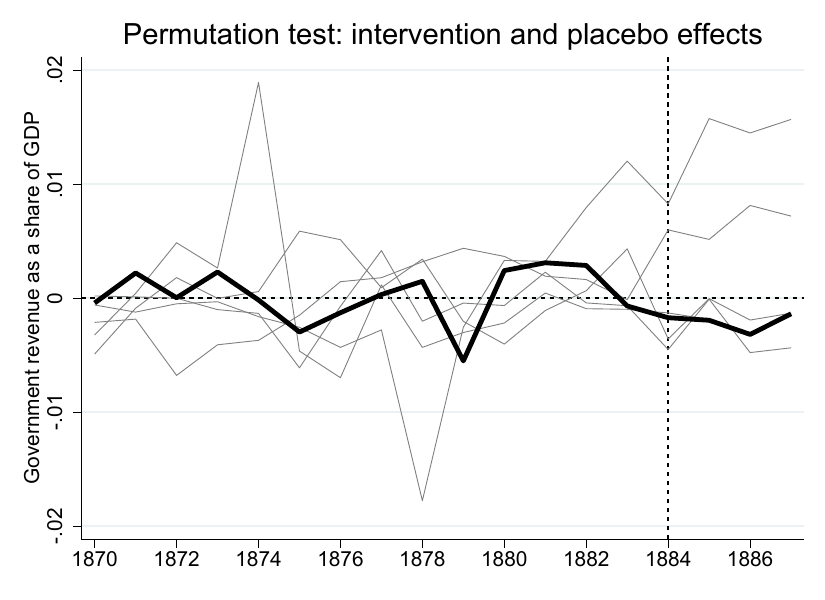}
        \includegraphics[width=0.48\textwidth]{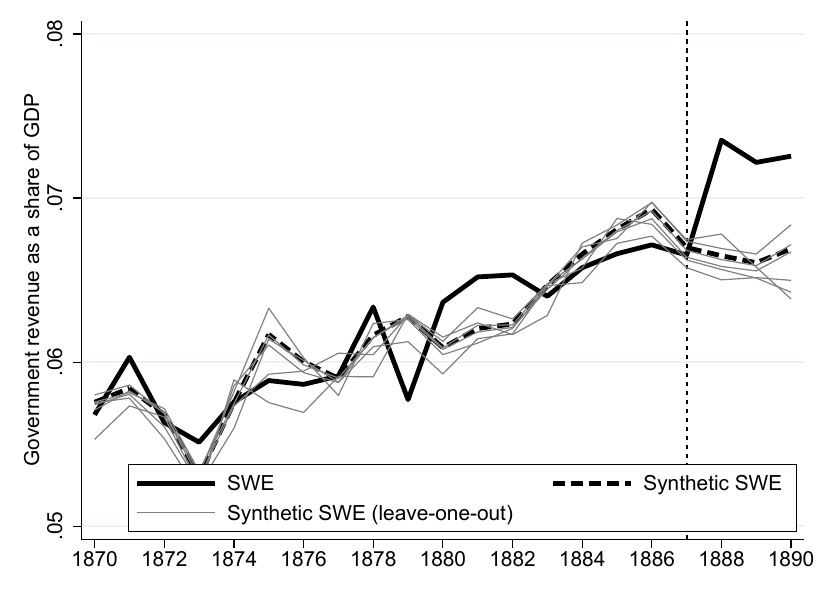}
	\end{center}
	\footnotesize{\textit{Notes:} The left panel shows the results for government revenue for the placebo treatment in 1884 and excludes countries for which the pre-treatment MSPE is at least 10 times larger than Sweden's pre-treatment MSPE. The right panel shows the Swedish counterfactuals for government revenue iteratively excluding each country in the donor pool with positive weights. The data are from the Jord\`{a}-Schularick-Taylor Macrohistory Database.}
	\label{fig: Government Revenue: Placebo Treatment in 1884 and Leave-one-out Sensitivity Check}
\end{figure}

First, following \citet{abadie2015comparative}, we backdate the treatment and consider a placebo change of government in the previous election year (1884). A significant effect of this placebo treatment would threaten the credibility of our findings. The results from the permutation inference procedure do not indicate an effect of the placebo treatment on government revenue (left panel of Figure \ref{fig: Government Revenue: Placebo Treatment in 1884 and Leave-one-out Sensitivity Check}). The ratio of post-treatment to pre-treatment RMSPE for Sweden is smaller than one and only the sixth highest among all countries. 

Second, we investigate whether an influential control country drives the estimated effect of protectionism on government revenue. Following \citet{abadie2015comparative}, we re-estimate the causal effect by iteratively excluding from the donor pool countries with a positive SC weight. The right panel of Figure \ref{fig: Government Revenue: Placebo Treatment in 1884 and Leave-one-out Sensitivity Check} shows the results. We find that the effect of protectionism on government revenue is not driven by any individual control countries.

\begin{figure}[H]
	\caption{Robustness: excluding countries with $>$10\% exports to Sweden in 1887}
	\begin{center}
		\includegraphics[width=0.39\textwidth,trim = {0 0cm 0 1.5cm}]{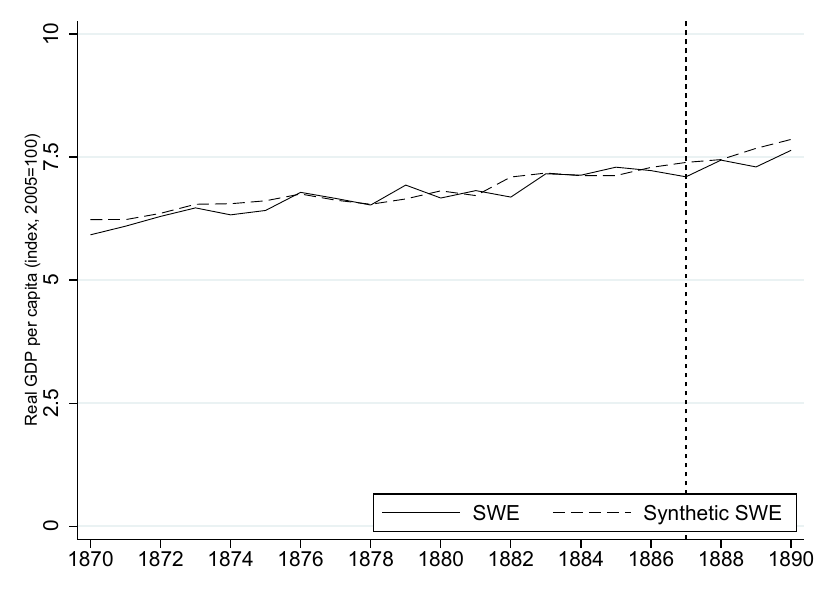}
		\includegraphics[width=0.39\textwidth,trim = {0 0cm 0 1.5cm}]{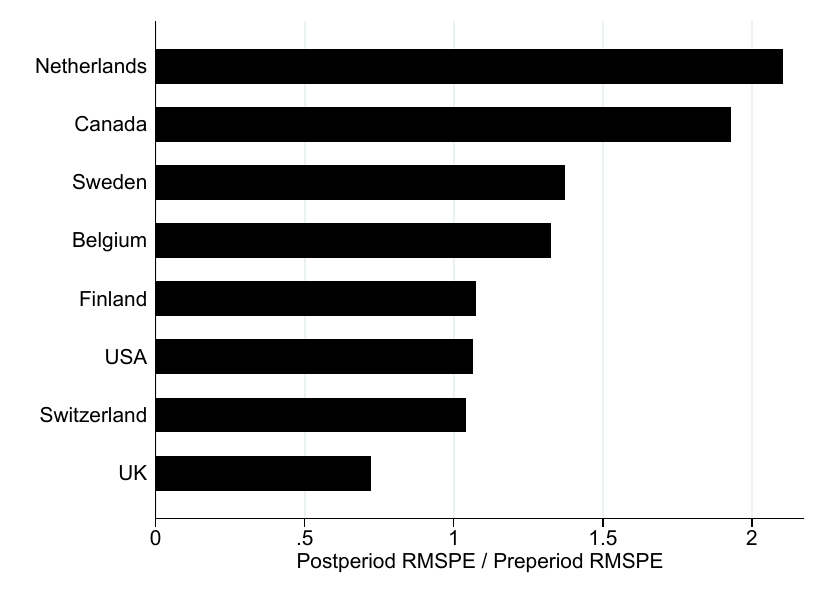}
        \includegraphics[width=0.39\textwidth,trim = {0 0cm 0 0cm}]{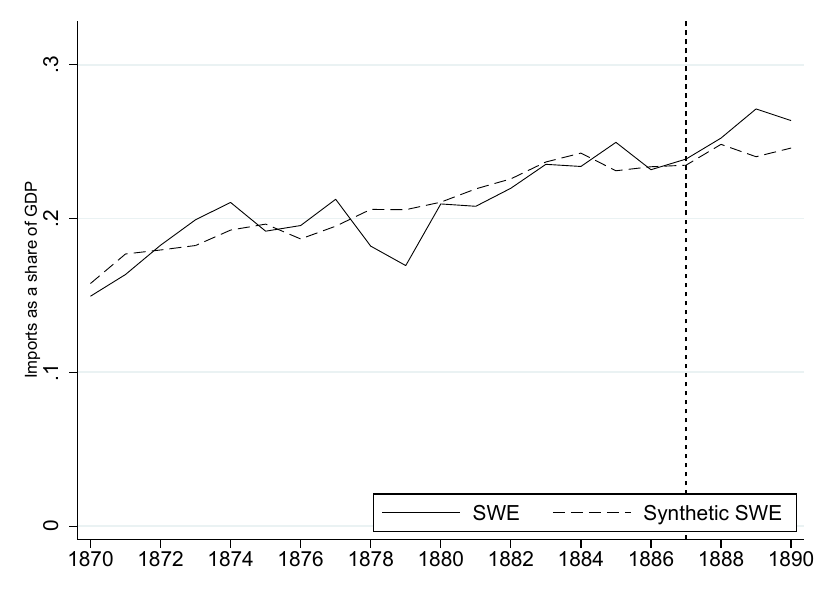}
        \includegraphics[width=0.39\textwidth,trim = {0 0cm 0 0cm}]{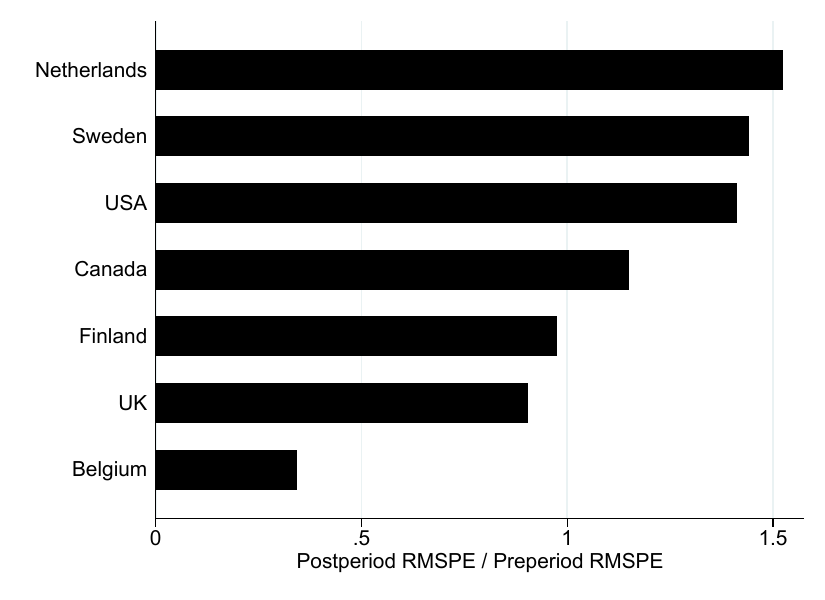}
        \includegraphics[width=0.39\textwidth,trim = {0 0cm 0 0cm}]{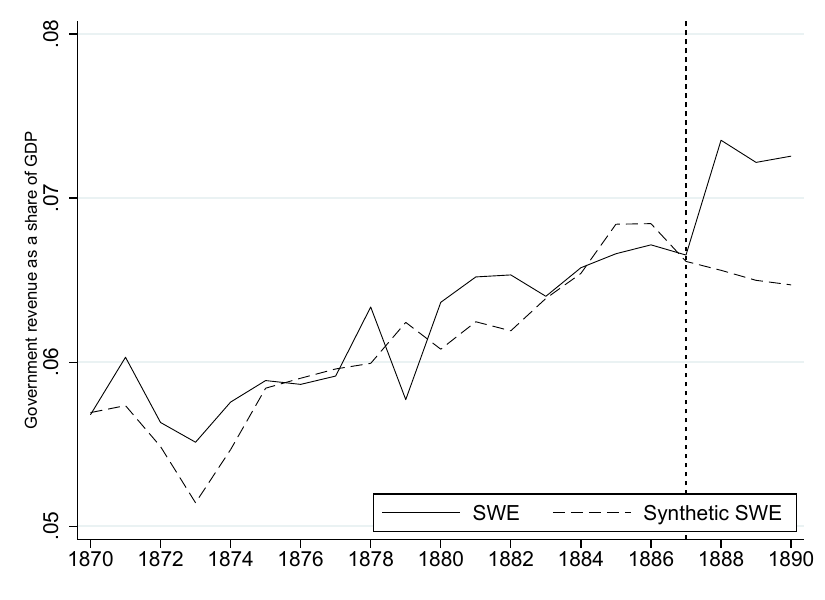}
        \includegraphics[width=0.39\textwidth,trim = {0 0cm 0 0cm}]{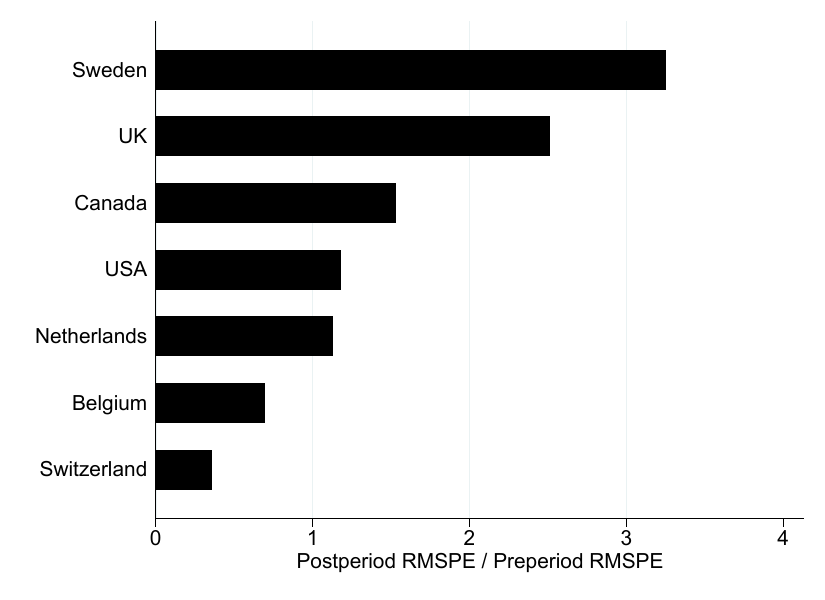}
        \includegraphics[width=0.39\textwidth,trim = {0 1cm 0 0cm}]{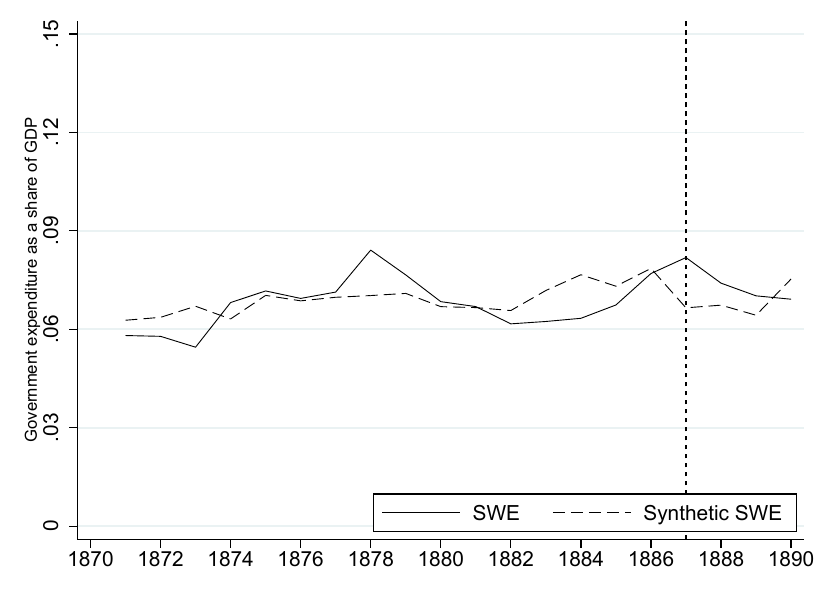}
        \includegraphics[width=0.39\textwidth,trim = {0 1cm 0 0cm}]{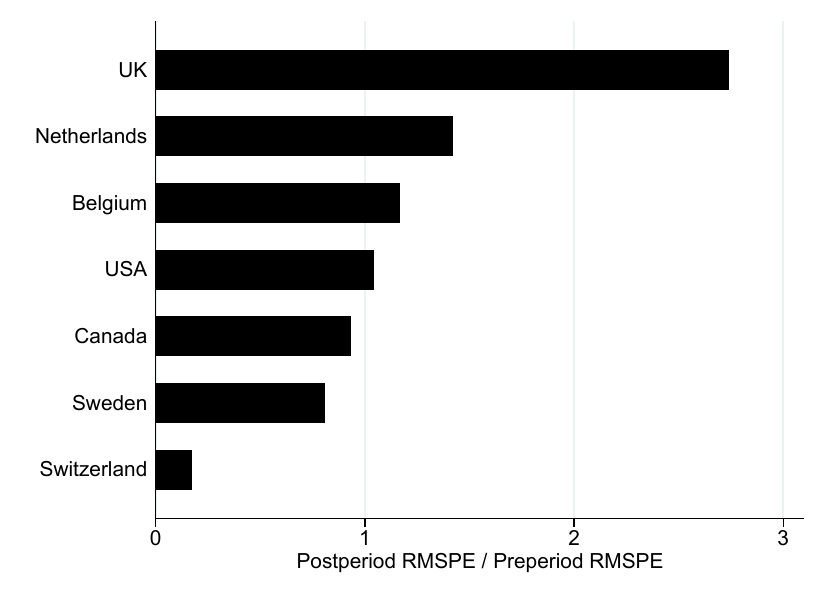}
	\end{center}
	\footnotesize{\textit{Notes:} We exclude Denmark and Norway from the original donor pool. The data are from the Jord\`{a}-Schularick-Taylor Macrohistory Database.}
	\label{fig: Robustness: Excluding high gravity countries}
\end{figure}

Third, we examine the robustness of our results to potential spillover effects from Sweden's tariff policy on its trading partners. We exclude from the donor pool countries exporting more than 10\% of their total exports to Sweden in 1887. There are two such countries: Denmark and Norway. Figure \ref{fig: Robustness: Excluding high gravity countries} shows that restricting the donor pool does not change the results. 

\begin{figure}[H]
	\caption{Robustness: European countries only}
	\begin{center}
		\includegraphics[width=0.39\textwidth,trim = {0 0cm 0 1.5cm}]{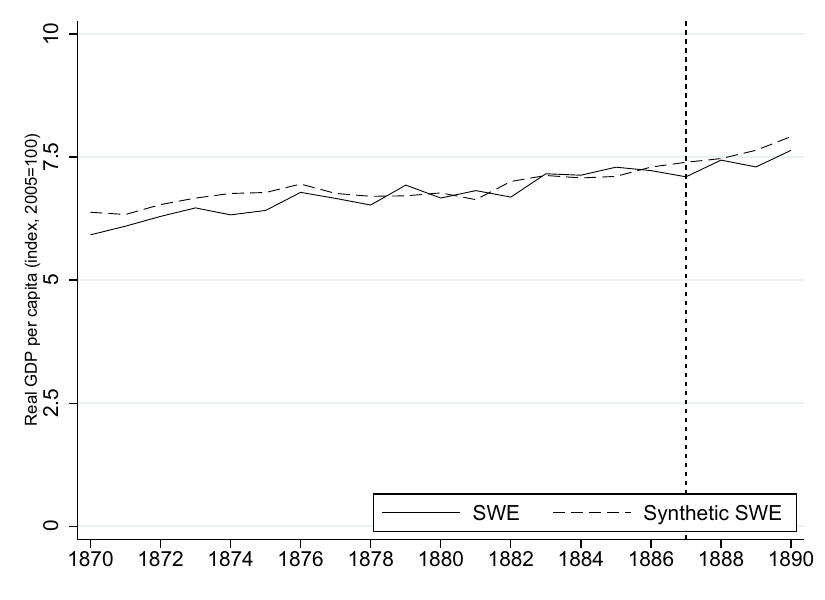}
		\includegraphics[width=0.39\textwidth,trim = {0 0cm 0 1.5cm}]{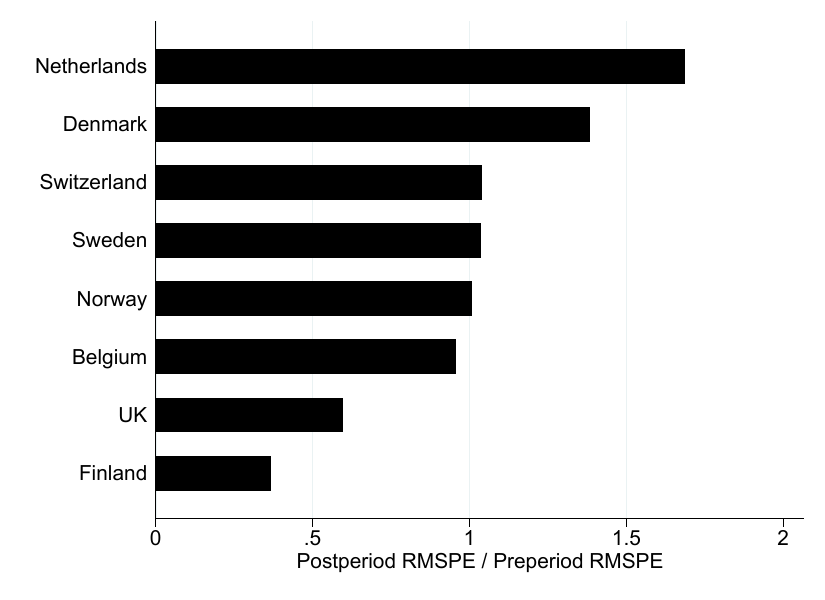}
        \includegraphics[width=0.39\textwidth]{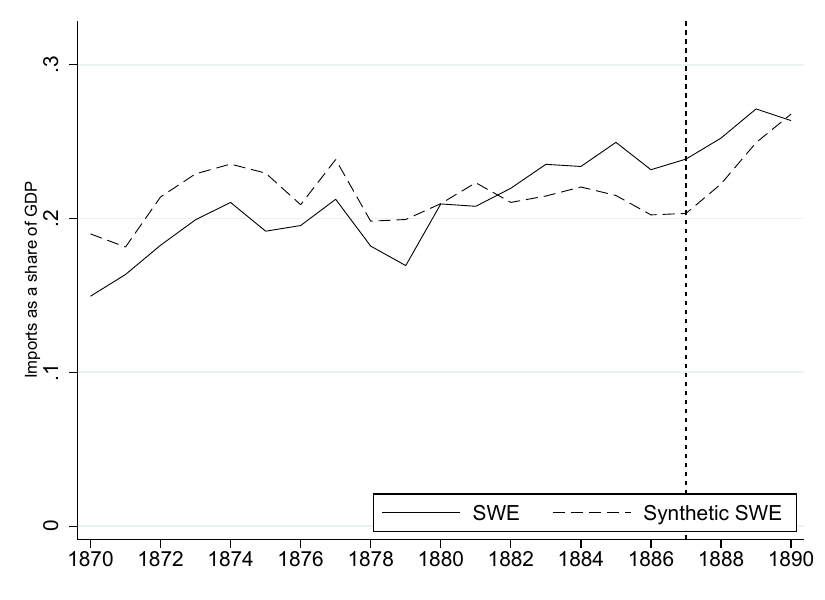}
        \includegraphics[width=0.39\textwidth]{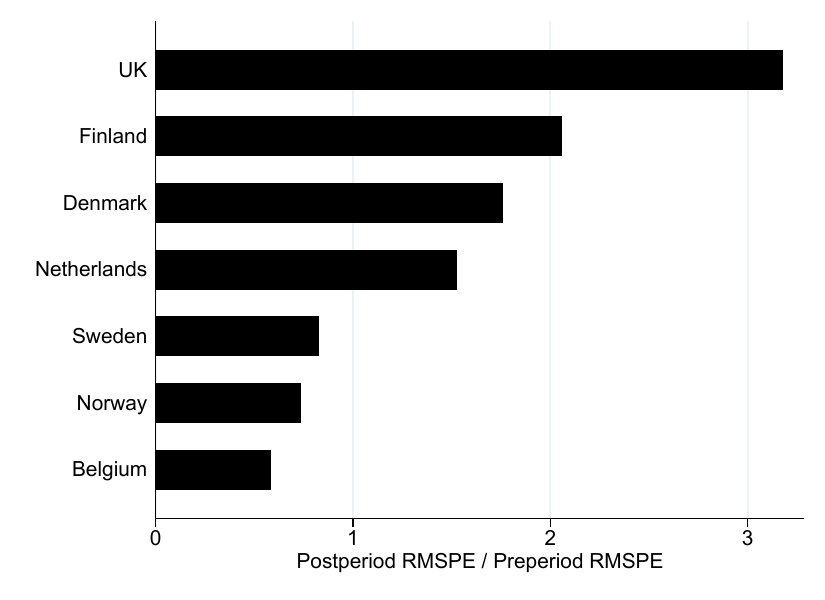}
        \includegraphics[width=0.39\textwidth]{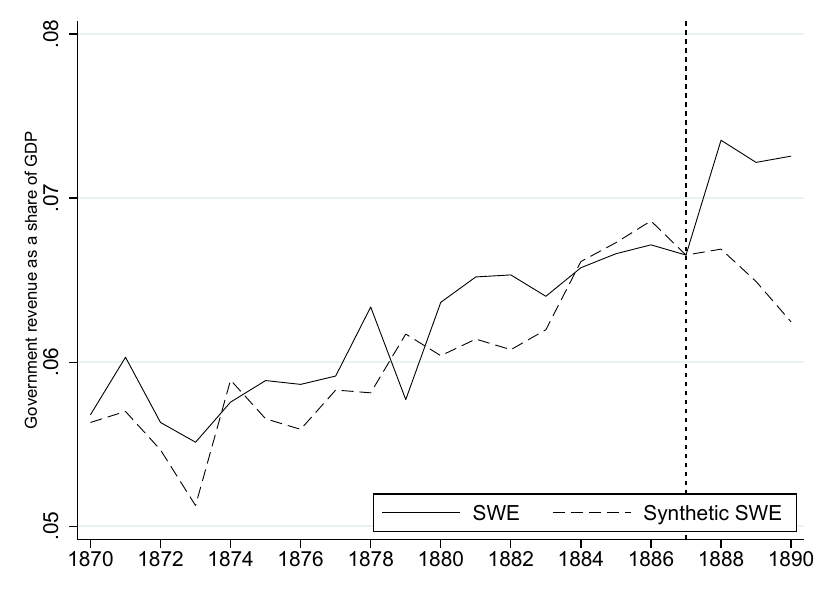}
        \includegraphics[width=0.39\textwidth]{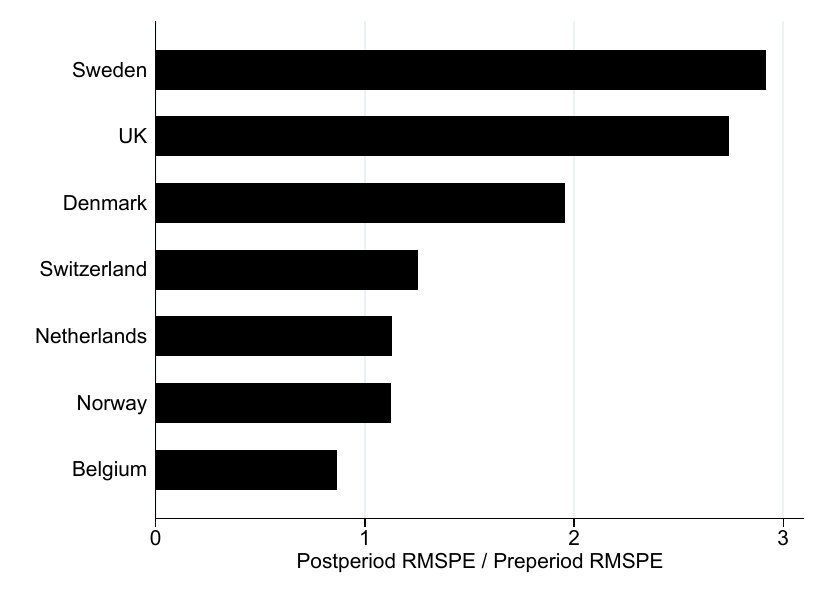}
        \includegraphics[width=0.39\textwidth,trim = {0 1cm 0 0cm}]{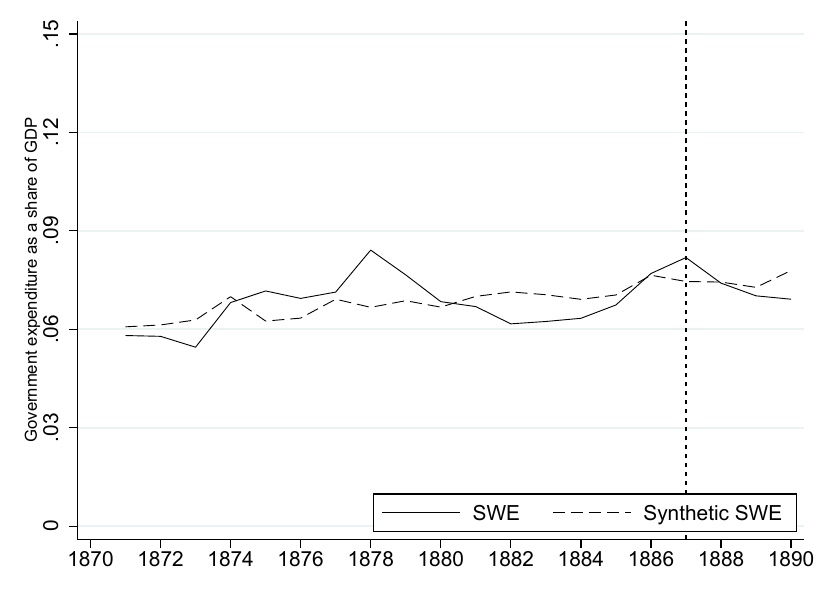}
        \includegraphics[width=0.39\textwidth,trim = {0 1cm 0 0cm}]{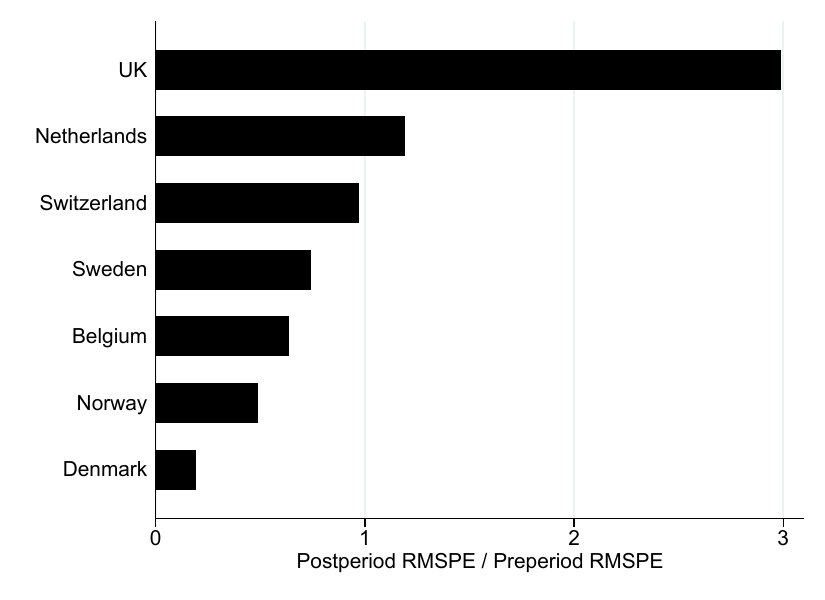}
	\end{center}
	\footnotesize{\textit{Notes:} We exclude Canada and the United States from the original donor pool. The data are from the Jord\`{a}-Schularick-Taylor Macrohistory Database.}
	\label{fig: Robustness: European countries only}
\end{figure}

Fourth, we restrict the donor pool to European countries. Average tariffs between 1870 and 1890 were substantially higher in the labor-scarce, land-abundant United States and Canada than in the European countries, and the institutional settings were different as well \citep{Irwin2002}. Figure \ref{fig: Robustness: European countries only} shows that excluding Canada and the United States from the donor pool does not change the inferences.

Finally, we apply the conformal inference procedure of \citet{chernozhukov2020exact} to test the null hypothesis that the protectionist policies had no effect on government revenue. This approach computes $p$-values by permuting SC residuals. We employ the two types of permutations recommended by \citet{chernozhukov2020exact}: iid and moving block permutations. The $p$-values are 0.01 for the iid permutations and 0.05 for the moving block permutations (the smallest possible $p$-value given the number of time periods). Thus, our results are robust to the choice of the inference procedure.

\section{Conclusion}
Previous studies did not explain the `tariff-growth paradox' in the first era of globalization: protectionism has been shown to decrease economic growth in the 20th and 21st century, but tariffs and growth were positively correlated in the late 19th century and the early 20th century. We provide causal evidence on how protectionist policies influenced short-run economic growth in the late 19th century. We exploit an exogenous shock, unique in Sweden's history, that changed the parliamentary majority from free-trade to protectionist. The new protectionist government increased tariffs.

Using the SC method, we do not find evidence suggesting that the protectionist policies influenced short-run economic growth in late 19th century Sweden. An interesting question is why. The results suggest that the increased tariffs did not deter Sweden's trading partners from exporting goods to Sweden. The protectionist government increased revenue but refrained from stimulating the economy in the short-run by increasing government expenditure. Instead, it used the increased government revenue to balance the budget. Protectionist policies may increase government revenue without jeopardizing economic growth when the tax base for other taxes and size of government are small.

More causal evidence is needed to better understand the `tariff-growth paradox' in the first era of globalization. The short-run effects of protectionism are likely to be context-specific \citep{eichengreen2019}. Empirical strategies to identify causal effects should also be employed to examine how individual tariffs (e.g., agricultural tariffs, industrial tariffs) influenced government revenue and growth \citep[e.g.,][]{lehmann2011restat}. 

Changes in import tariffs may also influence growth through firm productivity  \citep[e.g.,][]{pavcnik2002, muendler2004, amitikonings2007, fernandes2007, topalovakhandelwal2011, chensteinwender2021}. Future studies should therefore examine how protectionism in the late 19th century influenced firms' productivity and TFP. One avenue would be employing data at the firm level to investigate trade-policy shocks as scholars did for the late 20th and the early 21st century \citep[see][for a survey]{shusteinwender2019}. Another avenue would be extending historical data on TFP that is currently only available since the year 1890 \citep{bergeaud2016productivity}.

We examine how protectionism influences short-run economic growth 
Our post-treatment period encompasses three years. A relatively short post-treatment period is well-suited for our purpose because other confounding events after treatment are unlikely to bias our results. However, our research design is not suitable to identify and estimate the long-run effects of protectionism on economic growth, and future research should investigate such long-run effects in the first era of globalization.

\newpage

\begin{singlespace}
\interlinepenalty 10000
\setlength{\bibsep}{1pt}
\bibliographystyle{apalike}
\bibliography{sweden_bibliography.bib}
\end{singlespace}

\newpage

\appendix

\setcounter{page}{1} 

\begin{center}
    \LARGE{Appendix to \emph{Protectionism and economic growth: Causal evidence from the first era of globalization}}
    
    \vspace{1cm}
    
    \large{Niklas Potrafke \qquad Fabian Ruthardt \qquad Kaspar W\"uthrich}
    
    \vspace{1cm}
    
\end{center}

\spacingset{1.10} 

\startcontents[sections]
\printcontents[sections]{l}{1}{\setcounter{tocdepth}{1}}

\spacingset{1.25} 

\newpage

\section{1887 fall election results for the electoral district of Stockholm}
\label{sec: Election results Stockholm}

\begin{table}[H]
\centering
\footnotesize
\caption{Number of votes for free-trade, protectionist and independent candidates}

\def\sym#1{\ifmmode^{#1}\else\(^{#1}\)\fi}
\begin{tabular}{>{\color{Blue}}l| >{\color{Blue}}c|>{\color{Maroon}}l| >{\color{Maroon}}c|>{\color{Black}}l| >{\color{Black}}c}

\toprule
\textcolor{Black}{Free-trader}&\textcolor{Black}{Votes}&\textcolor{Black}{Protectionist}&\textcolor{Black}{Votes}&\textcolor{Black}{Independent}&\textcolor{Black}{Votes}\\
\hline
Key&6,707&de Laval&2,954&Telander&1,856\\
Nordenski\"old&6,641&Widstr\"om&2,946&Crusebj\"orn&1,777\\
Taube&6,640&Billing&2,984&Morssing&1,699\\
Fock&6,640&Palmstierna&2,982\\
von Friesen&6,639&Werner&2,876\\
Wallden&6,637&Styffe&2,787\\
Loven&6,636&Carlsson, E. W.&2,776\\
Stackelberg&6,627&Lindmark&2,756\\
Abergsson&6,626&Svanberg&2,731\\
Grafstr\"om&6,620&Bergman&2,717\\
Beckmann&6,617&Bexelius&2,717\\
Siljestr\"om&6,614&Berndes&2,716\\
Hedin&6,591&Nystr\"om C&2,715\\
Larsson, A. P.&6,497&Cederschi\"old&2,710\\
Johansson&6,475&Beskow&2,708\\
Fredholm&6,466&H\"oglund, F.&2,691\\
H\"oglund, O. M.&6,420&Bj\"orek&2,688\\
Erikson, P. J. M.&6,389&Carlsson, A. V.&2,688\\
Larsson, Olof&6,197&Lund&2,687\\
Hammarlund&4,916&Berg, C. O.&2,649\\
Berg, F.&4,911&Wittrock&2,628\\
Gustafsson&4,866&Lyth&2,598\\
\bottomrule

\end{tabular}

\begin{center}{\parbox[b]{15cm}{\footnotesize \emph{Sources}: \citet{Aftonbladet_18870924} 

}} \end{center}

\end{table}


\section{The Swedish tariff debate}
\label{sec: The Swedish tariff debate}

Economists have been investigating and discussing the effect of the Swedish tariff increases on the economy since 1888. 

The first scientific contribution on the matter came shortly after WWI. Eli Heckscher and Arthur Montgomery examined the effects of the Swedish 19th-century tariff policy in a public investigation. The final report was published in 1924 and concluded that the increased tariff protection was probably negative for the Swedish economy because it supported mostly domestic market industries and not export industries \citep{Tull-ochtraktatkommitten1924}. They later diverted from their previous assessment and argued that the tariffs probably had only small effects on the economy \citep{heckscher1941svenskt} and that Sweden would have developed similarly without the tariff increases in 1888 and 1892 \citep{montgomery1966industrialismens}.

In a similar vein, \citet{jorberg1961growth} describes that the tariffs may have contributed to import substitution but that the overall effect is difficult to assess. \citet{jorberg1966naagra} argued that the domestic market may have benefited from the tariffs but that this influenced the Swedish industrialization process only to a small extent. He concluded that the tariffs probably did not have a significant effect on Swedish industrial growth.

\citet{hammarstrom1970stockholm} argued that the tariffs triggered an import substitution process, particularly in the customer goods industries. Imports of finished products decreased, and raw material imports increased.

Contrary to previous work, \citet{schon1989kapitalimport} concluded that Swedish tariffs increased economic growth. Tariffs primarily protected industries with long-term growth potential and contributed to Sweden's industrial development. 

\citet{Bohlin2005} constructed tariff indices for a large part of the Swedish economy using a sample of commodities between 1885 and 1914, and, similar to Hammarstr\"{o}m, emphasized that the tariffs caused import substitution. Import penetration decreased significantly for goods subject to the tariffs of the late 1880s. 

\begin{quote}
\emph{``Even if one measures the tariff rate in a more appropriate way one may of course
argue that the rate of protection was not `high', however it was apparently high
enough in the majority of cases to achieve its aim of deterring imports. It seems
obvious that the protectionist system had effects, good or bad, on individual
industries and thus also on Swedish economic development in general.''} \\
--- \citep[][p.25]{Bohlin2005}
\end{quote}

More recently, \citet{Haeggqvist2018} contributed to the tariff debate by investigating the link between customs revenue and government activity. The Swedish trade liberalization initially forced a switch in the fiscal structure of tariffs towards consumption goods with low demand elasticity. After 1888, tariffs on agricultural and capital goods became more fiscally relevant.

\begin{quote}
\emph{``This development took place during a critical time when customs revenue as share of total government revenue really took off and came to be the single most important tax receipt. Trade policy hence came to be a key driver of nineteenth century fiscal development in Sweden.''} \\
--- \citep[][p.16]{Haeggqvist2018}
\end{quote}

The most comprehensive analysis of the topic so far was conducted by \citet{Persarvet_2019}. In his encompassing work, he concludes:

\begin{quote}
\emph{``Foreign trade and growth increased rapidly, the later more so after a protectionist trade policy was put in place in the late 1880s and 1890s.''} \\
--- \citep[][p.180]{Persarvet_2019}
\end{quote}

\begin{quote}
\emph{``In the end, the tariff protection thus probably had a limited impact on the
overall development of the aggregate productivity growth of the Swedish economy. Although it might have increased in the short term due to labor shifts, this effect was most likely small.''} \\
--- \citep[][p.184]{Persarvet_2019}
\end{quote}

We contribute to this longstanding debate by providing causal evidence on how the tariff increases influenced economic outcomes.

\section{Swedish fiscal policies in the first era of globalization}
\label{sec: Swedish fiscal policies in the first era of globalization}


\subsection{Government revenue}
Since 1870 Sweden's government revenue as a share of GDP remained relatively stable at around 6\%. In 1888, government revenue as a share of GDP increased substantially to well above 7\% and remained at this level throughout 1889 and 1890.

\subsubsection{Customs revenue}
\begin{figure}[H]
	\caption{Customs revenue}
	\begin{center}
		\includegraphics[width=0.48\textwidth]{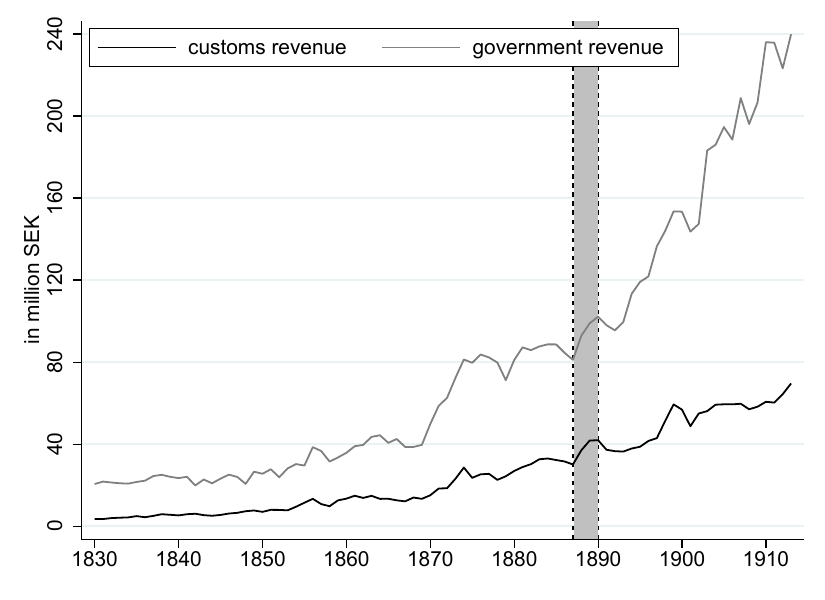}
		\includegraphics[width=0.48\textwidth]{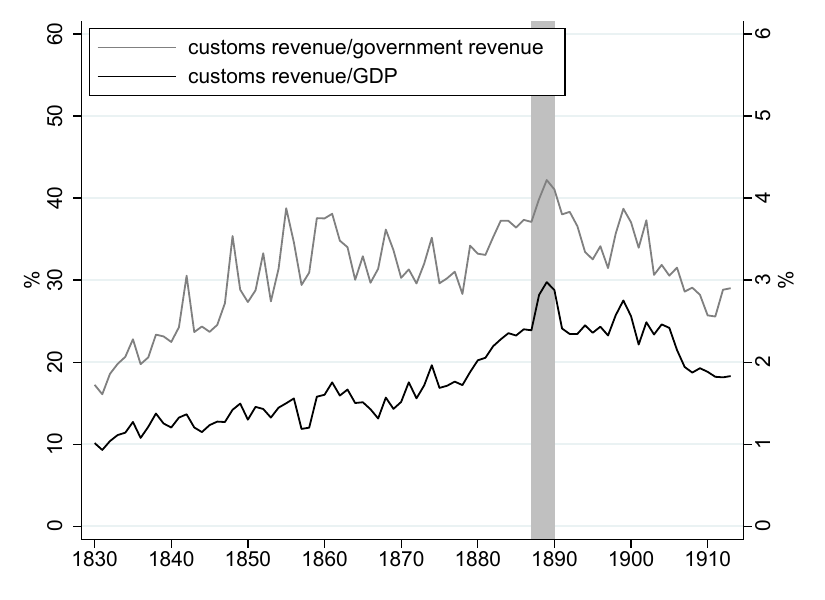}
	\end{center}
	\footnotesize{\textit{Notes:} The left panel shows the development of customs revenue and government revenue between 1830 and 1913. The right panel shows the development of customs revenue as a share of government revenue (left y-axis) and as a share of GDP (right y-axis) between 1830 and 1913. The relevant post-treatment period (1888--1890) is shaded in gray. The data are from \citet{Haeggqvist2018}.}
	\label{fig: customs revenue}
\end{figure}

Customs revenue was the single most important source of government revenue \citep{Haeggqvist2018}. Customs revenue as a share of total government revenue was 39.87\% in 1888, 42.20\% in 1889, and 41.06\% in 1890 --- the highest shares over the period from 1830 to 1913 (Figure \ref{fig: customs revenue}). Customs revenue increased by 6.94 million SEK or 23.06\% in 1888. Tariffs on grains accounted for more than half of the increase (4.16 million SEK), which changed the composition of customs revenue. While, in 1887, customs duties on agricultural products accounted for only 0.1\% of customs revenue, agricultural customs revenue as a share of total customs revenue increased to 14.7\% in 1888, 19.4\% in 1889, and peaked in 1890 (20.0\%). Customs revenue coming from industrial products was low; its share of total customs revenue was 2.9\% in 1887 and increased just slightly to 3.6\% in 1888, 3.8\% in 1889, and 3.7\% in 1890.\footnote{Customs duties on industrial products increased substantially in the 1890s. At the end of the decade, industrial customs revenue as a share of total customs revenue was above 10\%. Major increases occurred after the Cobden-Chevalier treaty expired in 1892: the share of total customs revenue coming from industrial products was 4.5\% in 1892, 6.5\% in 1893, 7.6\% in 1894, 8.2\% in 1895, 9.8\% in 1896, and 10.4\% in 1897 \citep{Haeggqvist2018}.}

\subsubsection{Taxation}
\begin{figure}[H]
	\caption{Tax revenue development}
	\begin{center}
		\includegraphics[width=0.6\textwidth]{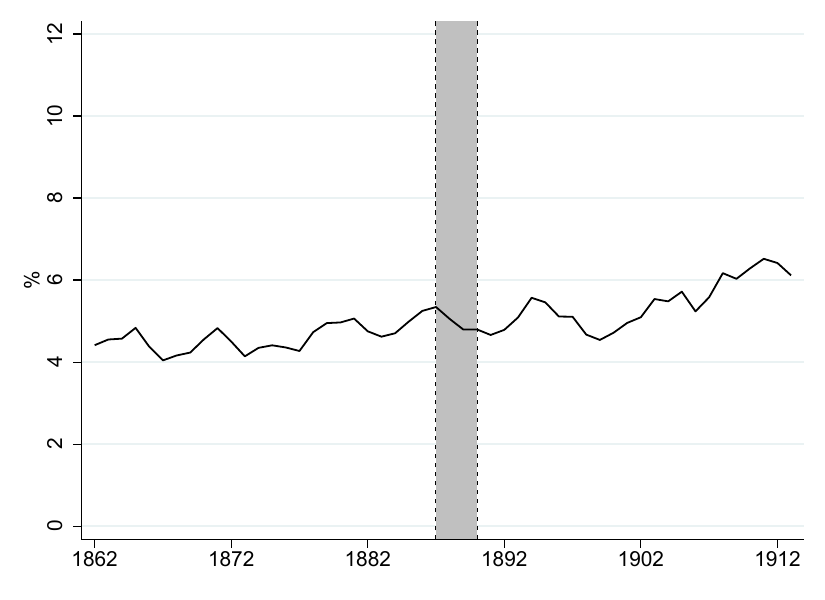}
	\end{center}
	\footnotesize{\textit{Notes:} The figure the development of tax revenue as a share of GDP between 1862 and 1913. The relevant post-treatment period (1888--1890) is shaded in gray. The data are from \citet{HenreksonStenkula2015}.}
	\label{fig: tax revenue}
\end{figure}

We follow the editorial work of \citet{HenreksonStenkula2015} and examine the development of taxation in Sweden for six key aspects of the Swedish tax system: the taxation of labor income, capital income, consumption, inheritance and gifts, wealth, and real estate. The overall tax-to-GDP ratio, excluding customs revenue, changed little between 1862 and 1913 (Figure \ref{fig: tax revenue}). Tax revenue usually fluctuated between four and 6\% of GDP.

\begin{figure}[H]
	\caption{Tax rate and revenue development by tax type}
	\begin{center}
		\includegraphics[width=0.48\textwidth]{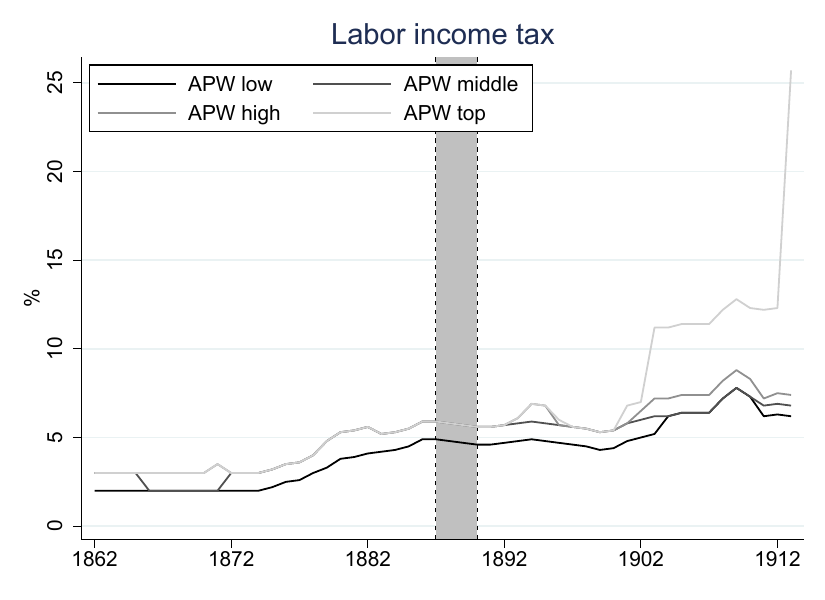}
		\includegraphics[width=0.48\textwidth]{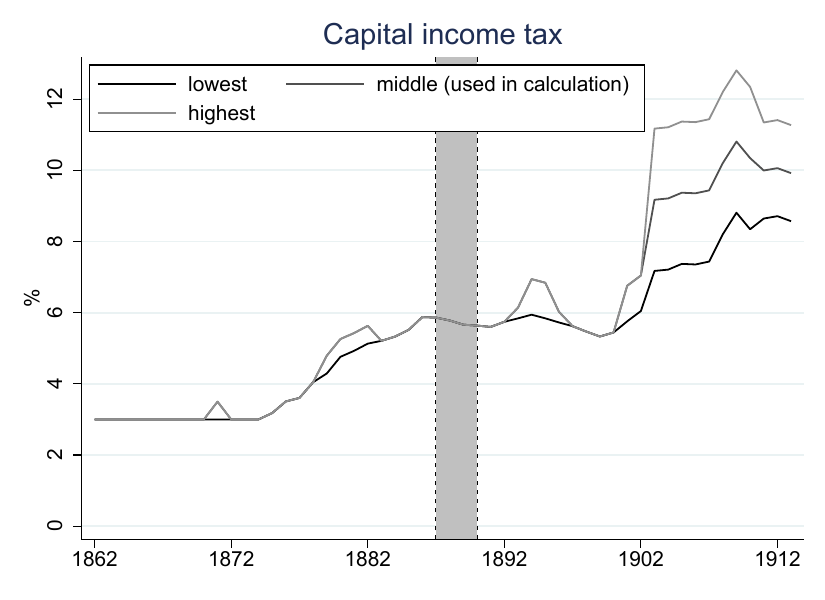}
		\includegraphics[width=0.48\textwidth]{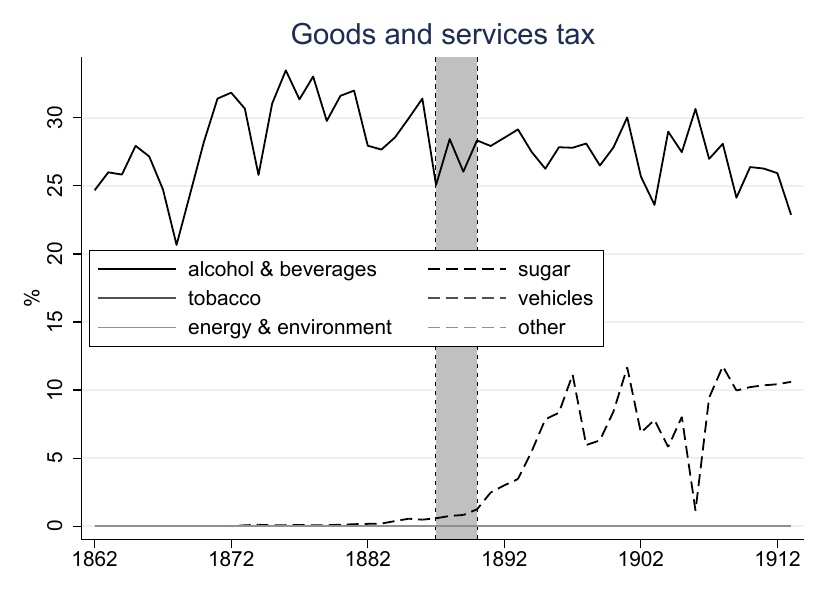}
		\includegraphics[width=0.48\textwidth]{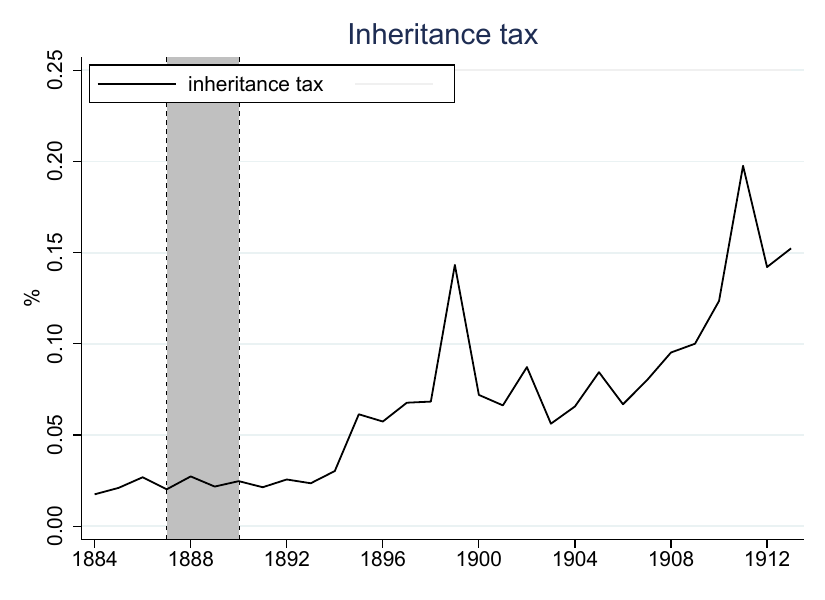}
	\end{center}
	\footnotesize{\textit{Notes:} The upper left panel shows the development of the marginal labor income tax rate as the sum of the national and local marginal labor income tax rates and social security contributions paid by employees. The upper right panel shows the development of the marginal effective tax rate on capital income for the highest and lowest statutory marginal corporate tax rate and the middle statutory marginal corporate tax rate used in the calculations by \citet{RietzJohanssonStenkula2015b}. The lower left panel shows the development of consumption taxes as a share of state tax revenue. The lower right panel shows the development of inheritance, estate, and gift tax revenue as a share of GDP. The relevant post-treatment period (1888--1890) is shaded in gray. The data are from \citet{HenreksonStenkula2015}.}
	\label{fig: tax revenue detail}
\end{figure}

Before WWI, major national income tax reforms were implemented in 1862, 1903, and 1911, none of which affected our pre-treatment period differently than our post-treatment period. The national tax level on labor income was normally set at 1\% but could be increased to 2\% if the ordinary appropriation taxes yielded insufficient revenue \citep{RietzJohanssonStenkula2015a}. However, in the years prior to and including 1887 and in our post-treatment period, the national marginal labor income tax rate remained constant. A local labor income tax, excise duties, and a national appropriation tax were also introduced in 1862 and 1863. The marginal local labor tax rate gradually increased from 2\% to 5\% at the end of the 19th century. Still, overall the marginal labor income tax rates remained low until WWI (Figure \ref{fig: tax revenue detail}, upper left panel).

Changes in capital income taxation affect the incentive to invest and thereby might influence GDP even in the short run. Again, the analysis of Swedish capital income taxation begins in 1862 with the introduction of a major central government tax system. The new state appropriation tax law taxed corporate profits in the same way and at the same rates as individual taxpayers' income (approximately 1\%, see previous paragraph\footnote{Initially, approximately 1\% of taxable profit was paid to the state, and approximately 2\% were paid to local governments. The state income tax was stable, but the local tax rate increased to approximately 5\% until 1900.}) \citep{RietzJohanssonStenkula2015b}. Based on \citet{kingfullerton1984}, \citet{RietzJohanssonStenkula2015b} calculate the marginal effective tax rate (METR) on capital income for an investment financed with new share issues, retained earnings, and debt. The METR was low between 1862 and 1913 and changed little in the 1880s (Figure \ref{fig: tax revenue detail}, upper right panel).

Consumption taxes were the second most important state revenue stream after customs duties between 1862 and 1913 \citep{Stenkula2015}. Consumption taxes accounted for 15\% to 20\% of total tax revenue and up to 40\% of state tax revenue until WWI. Alcohol-related taxes were the most important specific consumption tax, with revenue fluctuating between 20.7\% and 33.5\% of state tax revenue. Consumption taxes on sugar increased substantially during the 1890s from 1.24\% of state tax revenue in 1890 to around 10\% at the end of the century. Overall, revenue from consumption taxes fluctuated considerably, but while fluctuations were most pronounced in the 1860s and 1870s and sugar started to gain importance in the 1890s, the 1880s were a decade of comparatively stable consumption tax revenue (Figure \ref{fig: tax revenue detail}, lower left panel).

Finally, various duties and fees on estates, inheritances, and wills existed for small and parts of the tax base and population strata throughout the 18th century. In 1885, the modern Swedish inheritance taxation was introduced as a single tax (the 1884 Stamp Ordinance) \citep{RietzHenreksonWaldenstrom2015}. The actual tax was imposed on the lots received by the heirs. The income tax reform of 1861/62 also reduced the inheritance tax from around 3\% to a flat rate at 1\%. The 1884 Stamp Ordinance merged all previous variants of estate taxes (e.g., stamp duties and inheritance lot taxes) into a single tax in the form of a stamp on the total estate value. Direct heirs were taxed at a rate of 0.5\%, and other heirs were taxed at a rate of 0.6\%. \citet{RietzHenreksonWaldenstrom2015} find that revenue from the gift, inheritance, and estate taxes were never fiscally important when compared to personal income or wealth taxes, even though tax rates increased substantially in the 20th century. They suggest, that the inheritance tax was primarily supposed to reduce large intergenerational transfers at the top of the distribution. Accordingly, inheritance, estate, and gift tax revenue in Sweden was very low and accounted for only 0.017\% of GDP in 1884 and 0.030\% of GDP in 1894 (Figure \ref{fig: tax revenue detail}, lower right panel). Swedish wealth taxation was introduced in 1911 and is therefore not suitable for consideration in our study \citep{RietzHenrekson2015}.


\subsection{Government expenditure}
\subsubsection{Central government expenditure} 
\begin{figure}[H]
	\caption{Central government expenditure 1830--1913}
	\begin{center}
		\includegraphics[width=0.8\textwidth]{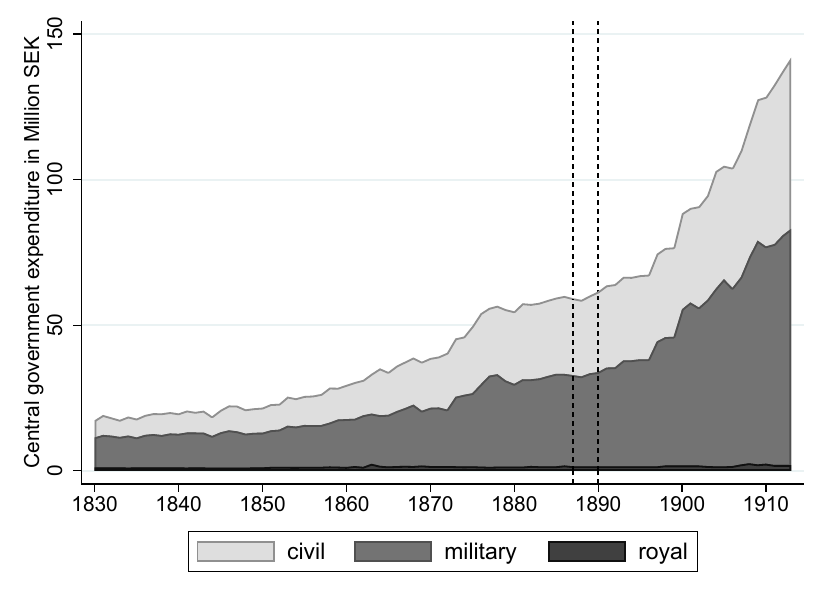}
	\end{center}
	\footnotesize{\textit{Notes:} The relevant post-treatment period (1888--1890) is indicated by the dashed vertical lines. The data are from  \citet{SchoenKrantz2015}.}
	\label{fig: Central government expenditure 1830--1913}
\end{figure}

Swedish central government expenditure increased at an accelerating pace from 1830--1913 (Figure \ref{fig: Central government expenditure 1830--1913}). However, the 1880s and early 1890s were marked by rather steady and moderate increases. In 1888, the increased central government revenue due to the increased customs revenue gave rise to financial desires across the parliamentary benches and the royal court \citep{Britannica1911}. On October 12, 1888, Oscar II\footnote{Oscar II was King of Sweden from 1872--1907.} declared at the Council of State that he wishes to spend the surplus from the increased customs revenue on insurance and pensions, the abolition of the land taxes, and lowering of the municipal taxes. However, the \textit{Riksdag} devoted the increased central government revenue to balance the budget \citep{Britannica1911}. Overall, the budget composition changed little after the majority in parliament changed \citep{SchoenKrantz2015}. Civil, military, and royal expenditure remained unchanged or increased just slightly, thereby following its slow but steady growth path.

\subsubsection{Local government expenditure} 

\begin{figure}[H]
	\caption{Local government expenditure 1830--1913}
	\begin{center}
		\includegraphics[width=0.8\textwidth]{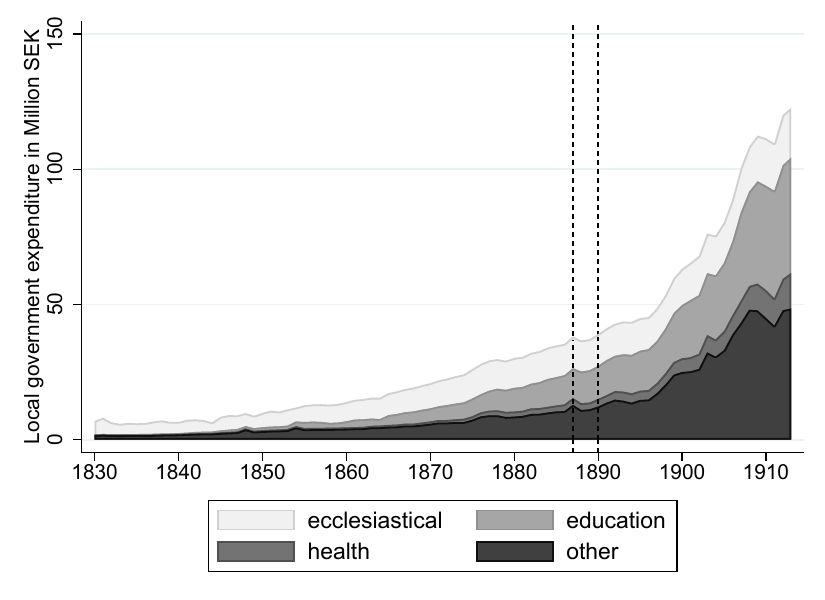}
	\end{center}
	\footnotesize{
	\textit{Notes:} The relevant post-treatment period (1888--1890) is indicated by the dashed vertical lines. The data are from  \citet{SchoenKrantz2015}.}
	\label{fig: Local government expenditure 1830--1913}
\end{figure}

Swedish local government expenditure developed similarly to Swedish central government expenditure, though being less volatile pre-1900 (Figure \ref{fig: Local government expenditure 1830--1913}). Local government expenditure ought to be unaffected by the majority change in the \textit{Riksdag} and the change in national government. If we observe an increase or decrease in local government expenditure after 1887, our identification strategy would likely be biased because local government expenditure potentially influences short-run macroeconomic outcomes. We do not observe that local government expenditure changed systematically after 1887. Indeed, \citet{SchoenKrantz2015} show that ecclesiastical expenditure hardly changes throughout the 1870s and 1880s whereas educational expenditure follows a steady growth path that accelerates at the turn of the century. Health expenditure remained a minor item on the local government expenditure list until the early 1900s.

\section{Composition of Swedish imports}

\subsection{Swedish imports by trading partner}
\label{sec: Imports by trade partner}

\begin{figure}[H]
	\caption{Swedish imports by trading partner: 1870 to 1890}
	\begin{center}
		\includegraphics[width=1\textwidth]{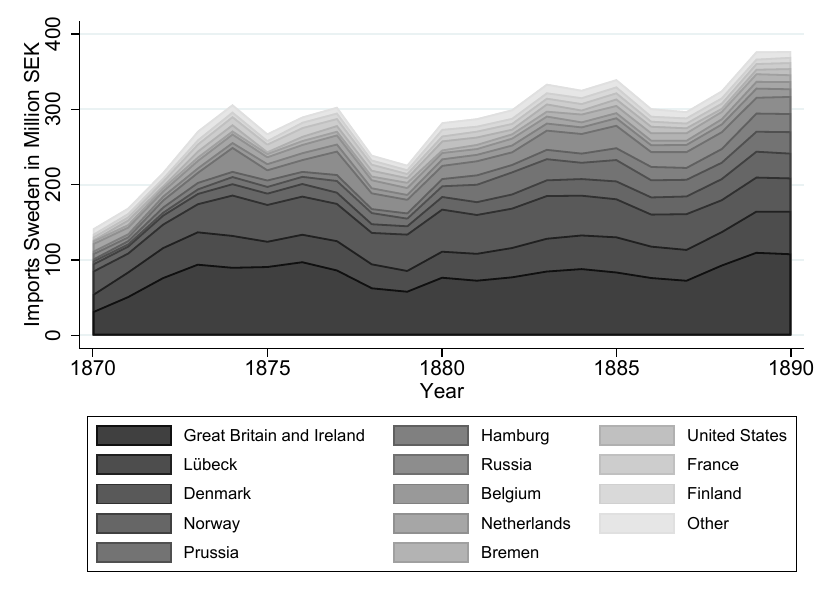}
	\end{center}
	\footnotesize{\textit{Notes:} The data are from the Swedish Board of Trade: Annual Statistics 1870 to 1890.}
	\label{fig: Swedish Imports from Trade Partners: 1870 to 1890}
\end{figure}

\clearpage

\begin{sidewaystable}
\caption{Swedish imports by trading partner: growth rates}
\resizebox{\textwidth}{!}{
\begin{tabular}{llllllllllllllllllllll}
Countries                   & 1871  & 1872  & 1873  & 1874  & 1875  & 1876  & 1877  & 1878  & 1879  & 1880 & 1881  & 1882  & 1883 & 1884  & 1885 & 1886  & 1887  & 1888  & 1889  & 1890  \\
\hline
Norway                      & -4\%  & 29\%  & 14\%  & 11\%  & 0\%   & 11\%  & -11\% & -23\% & -4\%  & 51\% & 1\%   & 11\%  & 11\% & 6\%   & 8\%  & -4\%  & 3\%   & 20\%  & 22\%  & -4\%  \\
Finland                     & 23\%  & -11\% & 40\%  & 10\%  & -24\% & 52\%  & 3\%   & -24\% & -8\%  & 73\% & -25\% & 3\%   & 6\%  & 9\%   & 0\%  & -16\% & 4\%   & 2\%   & -17\% & 20\%  \\
Russia                      & -29\% & 87\%  & 44\%  & 138\% & -57\% & 15\%  & 100\% & -33\% & -13\% & -3\% & 7\%   & -10\% & 51\% & 2\%   & 15\% & -34\% & 6\%   & 6\%   & -6\%  & 9\%   \\
Denmark                     & -19\% & 24\%  & 20\%  & 44\%  & -9\%  & 4\%   & -2\%  & -16\% & 15\%  & 16\% & -8\%  & 1\%   & 9\%  & -7\%  & -4\% & -16\% & 12\%  & -11\% & 7\%   & -3\%  \\
Prussia                     & -27\% & 34\%  & 60\%  & 57\%  & -7\%  & 7\%   & 64\%  & -5\%  & -31\% & 40\% & 63\%  & 27\%  & -4\% & -23\% & 31\% & -20\% & -3\%  & -2\%  & 22\%  & 8\%   \\
Luebeck    & 43\%  & 22\%  & 8\%  & -2\%  & -21\% & 9\%   & 7\%   & -18\% & -13\% & 25\%  & 3\%   & 9\%   & 12\% & 3\%   & 4\%  & -11\% & -2\%  & 8\%   & 23\%  & 4\%   \\
Hamburg                     & 70\%  & 58\%  & 24\%  & -20\% & 25\%  & -23\% & 20\%  & -33\% & 31\%  & 38\% & 31\%  & -21\% & 27\% & -2\%  & 29\% & 13\%  & -11\% & 15\%  & 32\%  & 0\%   \\
Bremen    & 0\%   & 18\%  & 9\%  & -9\%  & 6\%   & 22\%  & -5\%  & 1\%   & 41\%  & -51\% & 60\%  & 4\%   & 14\% & 59\%  & -8\% & 2\%   & 2\%   & -3\%  & 5\%   & -11\% \\
Netherlands                 & -13\% & -23\% & 45\%  & 14\%  & -19\% & 19\%  & 11\%  & -12\% & -14\% & -8\% & -9\%  & 19\%  & 2\%  & -29\% & 2\%  & -10\% & -9\%  & 26\%  & 29\%  & 8\%   \\
Belgium                     & 23\%  & 28\%  & 14\%  & 10\%  & 25\%  & 12\%  & 3\%   & -28\% & -4\%  & 35\% & -2\%  & 6\%   & 4\%  & -6\%  & 13\% & -9\%  & 5\%   & 7\%   & 19\%  & -17\% \\
UK and Ireland & 62\%  & 49\%  & 23\%  & -4\%  & 1\%   & 7\%   & -11\% & -27\% & -7\%  & 31\% & -5\%  & 6\%   & 10\% & 4\%   & -5\% & -9\%  & -5\%  & 27\%  & 18\%  & -2\%  \\
France                      & 1\%   & 43\%  & 58\%  & 15\%  & -6\%  & 15\%  & -20\% & -14\% & -32\% & 18\% & 20\%  & -2\%  & 0\%  & 12\%  & -5\% & -17\% & -8\%  & 9\%   & 13\%  & 4\%   \\
USA               & 450\% & -53\% & 179\% & 4\%   & -70\% & 126\% & 34\%  & 2\%   & -20\% & 95\% & -18\% & -37\% & 56\% & -44\% & 61\% & 0\%   & -24\% & -37\% & 41\%  & 39\%  \\
Other     & -2\%  & 7\%   & 50\% & -16\% & -8\%  & -21\% & 10\%  & -18\% & -6\%  & 48\%  & 11\%  & 15\%  & 0\%  & -14\% & -3\% & 4\%   & -17\% & 12\%  & 5\%   & -22\% \\
\hline
\hline
Total                       & 19\%  & 28\%  & 25\%  & 13\%  & -13\% & 8\%   & 4\%   & -21\% & -5\%  & 25\% & 2\%   & 4\%   & 11\% & -2\%  & 4\%  & -11\% & -1\%  & 9\%   & 16\%  & 0\%  
\end{tabular}}
\begin{center}{\parbox[b]{20.3cm}{\footnotesize{\textit{Notes:} The data are from the Swedish Board of Trade: Annual Statistics 1870 to 1890.}
}}\end{center}
\end{sidewaystable}

\clearpage

\subsection{Swedish imports by sector}
\label{sec: Swedish imports by sector}

\begin{figure}[H]
	\caption{Swedish imports by sector: 1870 to 1890}
	\begin{center}
		\includegraphics[width=1\textwidth]{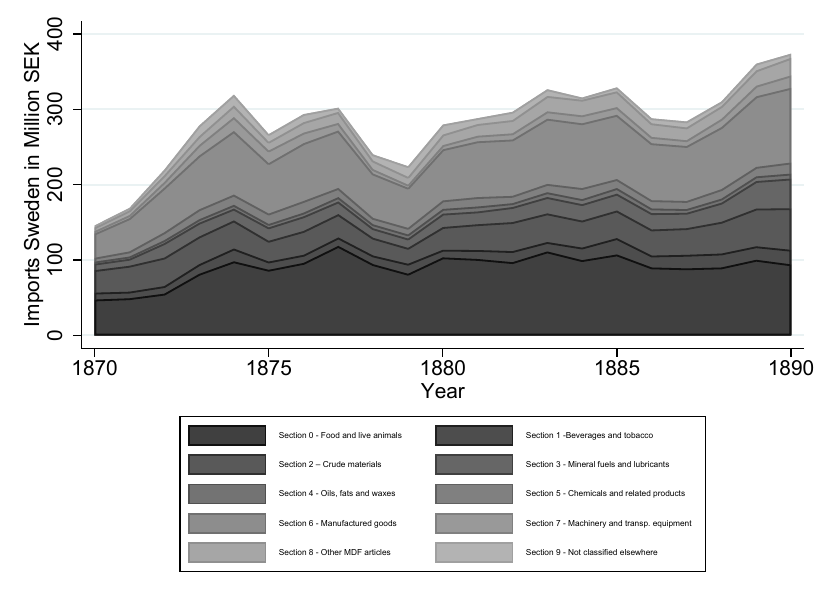}
	\end{center}
	\footnotesize{\textit{Notes:} The data are from \citet{Persarvet_2019}.}
	\label{fig: Swedish imports by sector: 1870 to 1890}
\end{figure}

\clearpage

\section{Synthetic control weights}
\label{sec: Weights}

\begin{table}[H]
\centering
\begin{small}
\caption{Synthetic control weights by outcome}
\begin{tabular}{lcccc}

\toprule
          & \multicolumn{1}{c}{\textbf{GDP}} & \multicolumn{1}{c}{\textbf{Imports}} 
          & \multicolumn{1}{c}{\textbf{Government}} & \multicolumn{1}{c}{\textbf{Government}} \\
          & \multicolumn{1}{c}{\textbf{}} & \multicolumn{1}{c}{\textbf{}} 
          & \multicolumn{1}{c}{\textbf{Revenue}} & \multicolumn{1}{c}{\textbf{Expenditure}}
          \\ \cmidrule{2-5}     \\
Belgium     & 0 & 0 & 0.128 & 0 \\
Canada      & 0 & 0 & 0.410 & 0 \\
Denmark     & 0.217 & 0.365 & 0.061 & 0.222  \\
Finland     & 0.436 & 0 &  $\cdot$   &  $\cdot$    \\
Netherlands & 0 & 0.102 & 0.126 & 0.229   \\
Norway      & 0.173 & 0 & 0 & 0.052   \\
Switzerland & 0 &  $\cdot$   & 0 & 0   \\
UK          & 0.003 & 0 & 0.231 & 0.314  \\
USA         & 0.170 & 0.532 & 0.045 & 0.182  \\
\\
\bottomrule
\end{tabular}
\label{tab: Weights}
\end{small}
\end{table}

\end{document}